\documentclass[a4paper,9pt]{article}
 
\usepackage{amsbsy,amssymb,amsmath,bm}
\usepackage{graphicx,color,epsfig,rotate}

\usepackage{rotating}

\newcommand{\be}{\begin{equation}}
\newcommand{\ee}{\end{equation}}
\newcommand{\Ket}[1]{\left|#1  \right>}

\usepackage{amsmath,amsfonts,amssymb,latexsym,times}
\usepackage{verbatim}
\usepackage{graphicx}
\usepackage{bm}
\usepackage{epstopdf}
\usepackage[dvipsnames]{xcolor} 
\usepackage{pstricks,pst-fill,pst-text,pst-node,pstricks-add}

\usepackage{psfrag, subfigure}

\usepackage{tikz}
\usetikzlibrary{calc}
\usetikzlibrary{chains}
\usetikzlibrary{scopes}
\usetikzlibrary{shapes.geometric}
\usetikzlibrary{trees}
\usetikzlibrary{topaths}
\usetikzlibrary{positioning}
\usetikzlibrary{patterns,decorations.pathreplacing}

\usepackage{multirow}

\newcommand{\cc}{{\mathbb P}\left(
\begin{tikzpicture}[baseline=8pt]
 \draw [fill] (0,0) circle(0.2ex);
 \draw [fill] (0,0.85) circle(0.2ex);
 \draw [thick] (0,0)--(0,0.85);
\end{tikzpicture}
\right)}

\newcommand{\ccsmall}{{\mathbb P}\left(
\begin{tikzpicture}[baseline=1pt]
 \draw [fill] (0,0) circle(0.2ex);
 \draw [fill] (0,0.25) circle(0.2ex);
 \draw [thick] (0,0)--(0,0.25);
\end{tikzpicture}
\right)}

\newcommand{\notccsmall}{{\mathbb P}\left(
\begin{tikzpicture}[baseline=1pt]
 \draw [fill] (0,0) circle(0.2ex);
 \draw [fill] (0,0.25) circle(0.2ex);
\end{tikzpicture}
\right)}

\newcommand{\Pdiff}{{\mathbb P}\left(
\begin{tikzpicture}
 \draw [fill] (0,0) circle(0.2ex);
 \draw [fill] (0.2,0) circle(0.2ex);
\end{tikzpicture}
\right)}

\newcommand{\Pa}{{\mathbb P}\left(
\begin{tikzpicture}[baseline=8pt]
 \draw [fill] (0,0) circle(0.2ex);
 \draw [fill] (0.2,0) circle(0.2ex);
 \draw [fill] (0,0.85) circle(0.2ex);
 \draw [fill] (0.2,0.85) circle(0.2ex);
 \draw [thick] (0,0)--(0,0.85);
  \draw [thick] (0.2,0)--(0.2,0.85);
\end{tikzpicture}
\right)}

\newcommand{\Pasmall}{{\mathbb P}\left(
\begin{tikzpicture}[baseline=1pt]
 \draw [fill] (0,0) circle(0.2ex);
 \draw [fill] (0.2,0) circle(0.2ex);
 \draw [fill] (0,0.25) circle(0.2ex);
 \draw [fill] (0.2,0.25) circle(0.2ex);
 \draw [thick] (0,0)--(0,0.25);
  \draw [thick] (0.2,0)--(0.2,0.25);
\end{tikzpicture}
\right)}

\newcommand{\Pb}{{\mathbb P}\left(
\begin{tikzpicture}[baseline=8pt]
 \draw [fill] (0,0) circle(0.2ex);
 \draw [fill] (0.2,0) circle(0.2ex);
 \draw [fill] (0,0.85) circle(0.2ex);
 \draw [fill] (0.2,0.85) circle(0.2ex);
 \draw [thick] (0,0)--(0,0.85);
\end{tikzpicture}
\right)}

\newcommand{\Pbsmall}{{\mathbb P}\left(
\begin{tikzpicture}[baseline=1pt]
 \draw [fill] (0,0) circle(0.2ex);
 \draw [fill] (0.2,0) circle(0.2ex);
 \draw [fill] (0,0.25) circle(0.2ex);
 \draw [fill] (0.2,0.25) circle(0.2ex);
 \draw [thick] (0,0)--(0,0.25);
\end{tikzpicture}
\right)}

\newcommand{\Pc}{{\mathbb P}\left(
\begin{tikzpicture}[baseline=8pt]
 \draw [fill] (0,0) circle(0.2ex);
 \draw [fill] (0.2,0) circle(0.2ex);
 \draw [fill] (0,0.85) circle(0.2ex);
 \draw [fill] (0.2,0.85) circle(0.2ex);
\end{tikzpicture}
\right)}

\newcommand{\Pcsmall}{{\mathbb P}\left(
\begin{tikzpicture}[baseline=1pt]
 \draw [fill] (0,0) circle(0.2ex);
 \draw [fill] (0.2,0) circle(0.2ex);
 \draw [fill] (0,0.25) circle(0.2ex);
 \draw [fill] (0.2,0.25) circle(0.2ex);
\end{tikzpicture}
\right)}

\newcommand{\PPa}{{\mathbb P}\left(
\begin{tikzpicture}[baseline=8pt]
 \draw [fill] (0,0) circle(0.2ex);
 \draw [fill] (0,0.2) circle(0.2ex);
 \draw [fill] (1,0) circle(0.2ex);
 \draw [fill] (1,0.2) circle(0.2ex);
 \draw [fill] (0.4,0.85) circle(0.2ex);
 \draw [fill] (0.6,0.85) circle(0.2ex);
 \draw [thick] (0,0)--(1,0)--(0.4,0.85)--cycle;
 \draw [thick] (0,0.2)--(1,0.2)--(0.6,0.85)--cycle;
 \end{tikzpicture}
\right)}

\newcommand{\PPb}{{\mathbb P}\left( 
\begin{tikzpicture}[baseline=8pt]
 \draw [fill] (0,0) circle(0.2ex);
 \draw [fill] (0,0.2) circle(0.2ex);
 \draw [fill] (1,0) circle(0.2ex);
 \draw [fill] (1,0.2) circle(0.2ex);
 \draw [fill] (0.4,0.85) circle(0.2ex);
 \draw [fill] (0.6,0.85) circle(0.2ex);
 \draw [thick] (0,0.2)--(0.4,0.85);
 \draw [thick] (0.6,0.85)--(1,0.2);
 \draw [thick] (0,0)--(1,0);
 \end{tikzpicture}
\right)}

\newcommand{\PPc}{{\mathbb P}\left(
\begin{tikzpicture}[baseline=8pt]
 \draw [fill] (0,0) circle(0.2ex);
 \draw [fill] (0,0.2) circle(0.2ex);
 \draw [fill] (1,0) circle(0.2ex);
 \draw [fill] (1,0.2) circle(0.2ex);
 \draw [fill] (0.4,0.85) circle(0.2ex);
 \draw [fill] (0.6,0.85) circle(0.2ex);
 \draw [thick] (0,0.2)--(1,0.2)--(0.6,0.85)--cycle;
 \draw [thick] (0,0)--(1,0);
 \end{tikzpicture}
\right)}

\newcommand{\PPcm}{{\mathbb P}\left(
\begin{tikzpicture}[baseline=8pt]
 \draw [thick] (0,0.2)--(1,0.2)--(0.6,0.85)--cycle;
 \draw [thick] (0,0)--(1,0);
 
 \draw [fill] (0,0) circle(0.2ex);
 \draw [fill] (0,0.2) circle(0.2ex);
 \draw [fill] (1,0) circle(0.2ex);
 \draw [fill] (1,0.2) circle(0.2ex);
 \draw [fill=red] (0.4,0.85) circle(0.2ex);
 \draw [fill=red] (0.6,0.85) circle(0.2ex) node[above,text width=0.05cm] {\color{red}$1$};
 \end{tikzpicture}
\right)}

\newcommand{\PPcmm}{{\mathbb P}\left(
\begin{tikzpicture}[baseline=8pt]
 \draw [thick] (0,0.2)--(1,0.2)--(0.6,0.85)--cycle;
 \draw [thick] (0,0)--(1,0);
 
 \draw [fill] (0,0) circle(0.2ex);
 \draw [fill] (0,0.2) circle(0.2ex);
 \draw [fill] (1,0) circle(0.2ex);
 \draw [fill] (1,0.2) circle(0.2ex);
 \draw [fill=red] (0.4,0.85) circle(0.2ex);
 \draw [fill=red] (0.6,0.85) circle(0.2ex) node[above,text width=0.05cm] {\color{red}$2$};
 \end{tikzpicture}
\right)}

\newcommand{\PPcmmm}{{\mathbb P}\left(
\begin{tikzpicture}[baseline=8pt]
 \draw [thick] (0,0.2)--(1,0.2)--(0.6,0.85)--cycle;
 \draw [thick] (0,0)--(1,0);
 
 \draw [fill] (0,0) circle(0.2ex);
 \draw [fill] (0,0.2) circle(0.2ex);
 \draw [fill] (1,0) circle(0.2ex);
 \draw [fill] (1,0.2) circle(0.2ex);
 \draw [fill=red] (0.4,0.85) circle(0.2ex);
 \draw [fill=red] (0.6,0.85) circle(0.2ex) node[above,text width=0.05cm] {\color{red}$3$};
 \end{tikzpicture}
\right)}

\newcommand{\PPd}{{\mathbb P}\left( 
\begin{tikzpicture}[baseline=8pt]
 \draw [fill] (0,0) circle(0.2ex);
 \draw [fill] (0,0.2) circle(0.2ex);
 \draw [fill] (1,0) circle(0.2ex);
 \draw [fill] (1,0.2) circle(0.2ex);
 \draw [fill] (0.4,0.85) circle(0.2ex);
 \draw [fill] (0.6,0.85) circle(0.2ex);
 \draw [thick] (0,0.2)--(1,0.2);
 \draw [thick] (0,0)--(1,0);
 \end{tikzpicture}
\right)}

\newcommand{\PPdm}{{\mathbb P}\left( 
\begin{tikzpicture}[baseline=8pt]
 \draw [fill] (0,0) circle(0.2ex);
 \draw [fill] (0,0.2) circle(0.2ex);
 \draw [fill] (1,0) circle(0.2ex);
 \draw [fill] (1,0.2) circle(0.2ex);
 \draw [fill=red] (0.4,0.85) circle(0.2ex);
 \draw [fill=red] (0.6,0.85) circle(0.2ex) node[above,text width=0.05cm] {\color{red}$1$};
 \draw [thick] (0,0.2)--(1,0.2);
 \draw [thick] (0,0)--(1,0);
 \end{tikzpicture}
\right)}

\newcommand{\PPdmm}{{\mathbb P}\left( 
\begin{tikzpicture}[baseline=8pt]
 \draw [fill] (0,0) circle(0.2ex);
 \draw [fill] (0,0.2) circle(0.2ex);
 \draw [fill] (1,0) circle(0.2ex);
 \draw [fill] (1,0.2) circle(0.2ex);
 \draw [fill=red] (0.4,0.85) circle(0.2ex);
 \draw [fill=red] (0.6,0.85) circle(0.2ex) node[above,text width=0.05cm] {\color{red}$2$};
 \draw [thick] (0,0.2)--(1,0.2);
 \draw [thick] (0,0)--(1,0);
 \end{tikzpicture}
\right)}

\newcommand{\PPdmmm}{{\mathbb P}\left( 
\begin{tikzpicture}[baseline=8pt]
 \draw [fill] (0,0) circle(0.2ex);
 \draw [fill] (0,0.2) circle(0.2ex);
 \draw [fill] (1,0) circle(0.2ex);
 \draw [fill] (1,0.2) circle(0.2ex);
 \draw [fill=red] (0.4,0.85) circle(0.2ex);
 \draw [fill=red] (0.6,0.85) circle(0.2ex) node[above,text width=0.05cm] {\color{red}$3$};
 \draw [thick] (0,0.2)--(1,0.2);
 \draw [thick] (0,0)--(1,0);
 \end{tikzpicture}
\right)}

\newcommand{\PPe}{{\mathbb P}\left(
\begin{tikzpicture}[baseline=8pt]
 \draw [fill] (0,0) circle(0.2ex);
 \draw [fill] (0,0.2) circle(0.2ex);
 \draw [fill] (1,0) circle(0.2ex);
 \draw [fill] (1,0.2) circle(0.2ex);
 \draw [fill] (0.4,0.85) circle(0.2ex);
 \draw [fill] (0.6,0.85) circle(0.2ex);
 \draw [thick] (0,0)--(1,0);
 \end{tikzpicture}
\right)}

\newcommand{\PPf}{{\mathbb P}\left(
\begin{tikzpicture}[baseline=8pt]
 \draw [fill] (0,0) circle(0.2ex);
 \draw [fill] (0,0.2) circle(0.2ex);
 \draw [fill] (1,0) circle(0.2ex);
 \draw [fill] (1,0.2) circle(0.2ex);
 \draw [fill] (0.4,0.85) circle(0.2ex);
 \draw [fill] (0.6,0.85) circle(0.2ex);
 \draw [thick] (0,0.2)--(0.4,0.85);
 \draw [thick] (0.6,0.85)--(1,0.2);
 \end{tikzpicture}
\right)}

\newcommand{\PPg}{{\mathbb P}\left(
\begin{tikzpicture}[baseline=8pt]
 \draw [fill] (0,0) circle(0.2ex);
 \draw [fill] (0,0.2) circle(0.2ex);
 \draw [fill] (1,0) circle(0.2ex);
 \draw [fill] (1,0.2) circle(0.2ex);
 \draw [fill] (0.4,0.85) circle(0.2ex);
 \draw [fill] (0.6,0.85) circle(0.2ex);
 \draw [thick] (0,0.2)--(1,0.2)--(0.6,0.85)--cycle;
 \end{tikzpicture}
\right)}

\newcommand{\PPh}{{\mathbb P}\left(
\begin{tikzpicture}[baseline=8pt]
 \draw [fill] (0,0) circle(0.2ex);
 \draw [fill] (0,0.2) circle(0.2ex);
 \draw [fill] (1,0) circle(0.2ex);
 \draw [fill] (1,0.2) circle(0.2ex);
 \draw [fill] (0.4,0.85) circle(0.2ex);
 \draw [fill] (0.6,0.85) circle(0.2ex);
 \end{tikzpicture}
\right)}

\newcommand{\PPPa}{{\mathbb P}\left(
\begin{tikzpicture}[baseline=8pt]
 \draw [fill] (0,0) circle(0.2ex);
 \draw [fill] (0.2,0) circle(0.2ex);
 \draw [fill] (0.4,0) circle(0.2ex); 
 \draw [fill] (0,0.85) circle(0.2ex);
 \draw [fill] (0.2,0.85) circle(0.2ex); 
 \draw [fill] (0.4,0.85) circle(0.2ex); 
 \draw [thick] (0,0)--(0,0.85);
 \draw [thick] (0.2,0)--(0.2,0.85);
 \draw [thick] (0.4,0)--(0.4,0.85);  
\end{tikzpicture}
\right)}

\newcommand{\PPPb}{{\mathbb P}\left(
\begin{tikzpicture}[baseline=8pt]
 \draw [fill] (0,0) circle(0.2ex);
 \draw [fill] (0.2,0) circle(0.2ex);
 \draw [fill] (0.4,0) circle(0.2ex); 
 \draw [fill] (0,0.85) circle(0.2ex);
 \draw [fill] (0.2,0.85) circle(0.2ex); 
 \draw [fill] (0.4,0.85) circle(0.2ex); 
 \draw [thick] (0,0)--(0,0.85);
 \draw [thick] (0.2,0)--(0.2,0.85);
\end{tikzpicture}
\right)}

\newcommand{\PPPc}{{\mathbb P}\left(
\begin{tikzpicture}[baseline=8pt]
 \draw [fill] (0,0) circle(0.2ex);
 \draw [fill] (0.2,0) circle(0.2ex);
 \draw [fill] (0.4,0) circle(0.2ex); 
 \draw [fill] (0,0.85) circle(0.2ex);
 \draw [fill] (0.2,0.85) circle(0.2ex); 
 \draw [fill] (0.4,0.85) circle(0.2ex); 
 \draw [thick] (0,0)--(0,0.85);
\end{tikzpicture}
\right)}

\newcommand{\PPPd}{{\mathbb P}\left(
\begin{tikzpicture}[baseline=8pt]
 \draw [fill] (0,0) circle(0.2ex);
 \draw [fill] (0.2,0) circle(0.2ex);
 \draw [fill] (0.4,0) circle(0.2ex); 
 \draw [fill] (0,0.85) circle(0.2ex);
 \draw [fill] (0.2,0.85) circle(0.2ex); 
 \draw [fill] (0.4,0.85) circle(0.2ex);  
\end{tikzpicture}
\right)}

\newcommand{\PkN}{{\mathbb P}\left(
\begin{tikzpicture}[baseline=8pt]
 \draw [fill] (0,0) circle(0.2ex);
 \draw [fill] (0.2,0) circle(0.2ex);
 \draw [fill] (0.85,0) circle(0.2ex); 
 \draw [fill] (0,0.85) circle(0.2ex);
 \draw [fill] (0.2,0.85) circle(0.2ex); 
 \draw [fill] (0.85,0.85) circle(0.2ex); 
 \draw [thick] (0,0)--(0,0.85);
 \draw [thick] (0.2,0)--(0.2,0.85);
 \draw [thick] (0.85,0)--(0.85,0.85);  
 \draw (0.2,0.425) node[right] {$\ldots$};
 \draw [fill] (1.05,0) circle(0.2ex);
 \draw [fill] (1.25,0) circle(0.2ex);
 \draw [fill] (1.90,0) circle(0.2ex); 
 \draw [fill] (1.05,0.85) circle(0.2ex);
 \draw [fill] (1.25,0.85) circle(0.2ex); 
 \draw [fill] (1.90,0.85) circle(0.2ex); 
 \draw (1.25,0.425) node[right] {$\ldots$};
 \draw [thick,decoration={brace,mirror},decorate] (0,-0.2)--(0.85,-0.2)
  node [pos=0.5,anchor=north,yshift=-0.05cm] {$k$};
 \draw [thick,decoration={brace},decorate] (0,1.05)--(1.90,1.05)
  node [pos=0.5,anchor=south,yshift=0.05cm] {$N$};
\end{tikzpicture}
\right)}

\begin{document}

\title{Operator content of the critical Potts model in $d$ dimensions \\ and logarithmic correlations}

\author{Romain Vasseur$^{1,2}$ and Jesper Lykke Jacobsen$^{3,4}$ \\
[2.0mm]
  ${}^1$ Department of Physics, University of California, Berkeley, Berkeley, CA 94720, USA \\
  ${}^2$ Materials Science Division, Lawrence Berkeley National Laboratory, Berkeley, CA 94720, USA \\
  ${}^3$LPTENS, 24 rue Lhomond, 75231 Paris, France \\
  ${}^4$ Universit\'e Pierre et Marie Curie, 4 place Jussieu, 75252 Paris, France}

\maketitle

\begin{abstract}

Using the symmetric group $S_Q$ symmetry of the $Q$-state Potts model, we classify the (scalar) operator content of its underlying field theory in arbitrary dimension. In addition to the usual identity, energy and magnetization operators, we find fields that generalize the $N$-cluster operators well-known in two dimensions, together with their subleading counterparts. We give the explicit form of all these operators -- up to non-universal constants -- both on the lattice and in the continuum limit for the Landau theory. We compute exactly their two- and three-point correlation functions on an arbitrary graph in terms of simple probabilities, and give the general form of these correlation functions in the continuum limit at the critical point. Specializing to integer values of the parameter $Q$, we argue that the analytic continuation of the $S_Q$ symmetry yields logarithmic correlations at the critical point in arbitrary dimension, thus implying a mixing of some scaling fields by the scale transformation generator. All these logarithmic correlation functions are given a clear geometrical meaning, which can be checked in numerical simulations. Several physical examples are discussed, including bond percolation, spanning trees and forests, resistor networks and the Ising model. We also briefly address the generalization of our approach to the $O(n)$ model.

\end{abstract}

\section{Introduction}

It is well-known that correlation functions in scale invariant theories exhibit power-law behavior. One of the early successes of two-dimensional conformal field theory (CFT) is to have classified the corresponding critical exponents in a class of conformally invariant theories known as unitary minimal models~\cite{BPZ,FQS,CardyMin}. However, many models of current interest in statistical physics do not possess unitarity. This is the case not only for models with quenched disorder, that are inherently non-unitary, but also for pure models of geometrical critical phenomena, such as percolation and the self-avoiding walk (see {\it e.g.}~\cite{Saleur87}), that arise as limits of generic non-unitary CFTs (that is, respectively, the $Q\to 1$ state Potts model, and the O($n$) loop model with $n \to 0$). A cognate situation arises in the minimal models themselves, like the Ising model, provided that the set of purely local operators ({\it e.g.}, spin and energy) is extended with operators describing the non-local correlations between geometrical objects, such as domain walls and cluster boundaries. One speaks in this case of extended minimal models \cite{PRZ}. In all these cases, the lack of unitarity makes possible the appearance of signs in front of the power-law terms. When further two or more critical exponents coincide, or differ by an integer, the power laws may combine to produce a logarithmic factor in some of the correlation functions \cite{LCFT2}. This can also be understood in terms of the resonance phenomenon that occurs in the Frobenius method for solving ordinary differential equations: when several roots of the indicial equation coincide, or differ by an integer, a logarithm is produced in one of the solutions.

Logarithmic conformal field theory (LCFT) is the framework within which logarithmic correlation functions of this type are properly described. More formally, a LCFT is a CFT whose symmetry algebra exhibits
reducible but indecomposable modules, that is, the operators displaying logarithmic behavior are linked up in a Jordan cell structure under the action of the dilatation operator. Following the early
pioneering results some twenty years ago~\cite{RozanskySaleur,Gurarie}, the field of LCFT has steadily gained momentum thanks to the combined efforts of theoretical physicists and mathematicians interested in
indecomposable representations of algebras (see {\it e.g.}~\cite{FlohrAML,[FGST3],MathieuRidout,GabRun,GabRunW2,GabRunW}).
While it was realized from early on that any non-trivial%
\footnote{E.g., excluding the theory consisting of just the identity operator, as well as theories that are direct sums of non-LCFTs with total a central charge that adds up to zero.}
CFT with $c=0$ is necessarily logarithmic~\cite{Gurarie2}, further insight can be gained by insisting
that physical operators must transform irreducibly under any additional discrete symmetry present in the model~\cite{CardyReplica}.
This applies in particular to
percolation and disordered models, where the relevant symmetries are respectively that of the symmetric group
$S_Q$ acting on Potts spin labels, and the $S_M$ replica symmetry. 
In both cases the validity of some of the arguments given is not limited to two dimensions \cite{LCFT2}.

A major step forward for the study of two-dimensional LCFT materialized with the understanding that indecomposability in the continuum limit ({\it i.e.}, in the representation theory of the
Virasoro algebra) is preceded by indecomposability in the corresponding finite-size lattice models~\cite{PRZ,RS07} ({\it i.e.}, in the Temperley-Lieb algebra and its boundary extensions \cite{GJSV13} which can also be interpreted geometrically in terms of loops \cite{JS08a,JS08b,DJS09}). As a result, many aspects of the  boundary versions of the extended minimal model LCFT have been understood in details (see~\cite{LCFT3} for a recent review). The study of the corresponding bulk theories is however impeded by the appearance of dauntingly
complicated representations (involving, in particular, Jordan cells or arbitrarily high rank), and accordingly progress has essentially been limited to a few sporadic cases, such as
the $GL(1|1)$ and $SL(2|1)$ spin chains that describe, in the continuum limit, respectively $c=-2$ symplectic fermions \cite{Symplecticfermions} and $c=0$ percolation LCFTs~\cite{VGJS,GRS1,GRS2,GRS3,GRSV}. 

In this paper we present a detailed study of the $Q$-state Potts model in $d>2$ dimensions from the bulk LCFT perspective. We construct a class of operators $t^{(k,N)}$, rank-$k$ tensors acting
on $N$ spins, that are irreducible under the $S_Q$ symmetry for generic $Q$. When $Q$ tends to a non-negative integer these operators involve divergent terms and hence must mix among themselves
in order to provide physically well-defined operators. We interpret these operators geometrically in terms of their action on percolation clusters (for $Q=1$) and, more generally, on Fortuin-Kasteleyn clusters (for arbitrary $Q$). In particular, we show how the operator mixing leads to logarithms in correlation functions that can be defined in purely geometrical terms.

In a previous publication~\cite{VJSPercoLogs} we have illustrated, generalizing the ideas of Cardy~\cite{CardyReplica}, the general mechanism by exhibiting, in the case of percolation, the logarithmic behavior of a two-point function of operators acting on $N=2$ spins. The present paper extends this study to arbitrary two- and three-point functions of operators acting on any number $N$ of spins. We discuss in details the physical implications for spanning trees and forests ($Q \to 0$), percolation ($Q \to 1$) and the Ising model ($Q=2$), which are the cases that correspond to critical theories in higher dimensions ($d > 2$). However, we limit for the time being the investigations to operators that are fully symmetric in their tensor indices ({\it i.e.}, that are scalar operators in the sense of CFT). A systematic study of operators corresponding to arbitrary Young diagrams is underway and will be reported elsewhere.

It is useful to recall that already in 1970 did Polyakov use global conformal invariance to clarify the structure of two- and three-point functions of quasi-primary fields~\cite{Polyakov}. The results can be written
\begin{eqnarray}
 \left \langle \phi_1(x_1) \phi_2(x_2) \right \rangle &=& \frac{\delta_{\Delta_1,\Delta_2}}{|x_1-x_2|^{2 \Delta_1}} \,, \label{2ptPolyakov} \\
 \left \langle \phi_1(x_1) \phi_2(x_2) \phi_3(x_3) \right \rangle &=& \frac{C_{123}}{|x_1-x_2|^{\Delta_1+\Delta_2-\Delta_3} |x_2-x_3|^{\Delta_2+\Delta_3-\Delta_1} |x_3-x_1|^{\Delta_3+\Delta_1-\Delta_2}} \,. \label{3ptPolyakov}
\end{eqnarray}
While this makes explicit the structural form of the correlation functions, the scaling dimensions $\Delta_i$ and structure constants $C_{ijk}$ are left unspecified. Only much later
were these parameters determined for specific two-dimensional models \cite{BPZ,FQS,DO,ZZ}, and the values of $C_{ijk}$ for geometrical observables in the Potts model of the type considered in the present paper
remain under active investigation to this day~\cite{Zamo,Jacopo}. It seems presently out of reach to determine $\Delta_i$ and $C_{ink}$ for $d>2$.
The present work can be seen as the analogue of Polyakov's, for the case when there is invariance under both global conformal transformations and the discrete $S_Q$ symmetry
of the Potts model. In particular, we can again determine the structure of two- and three-point functions---now involving logarithms---but complete results including the values of 
scaling dimensions, structure constants, as well as certain universal prefactors (closely related to the so-called indecomposability parameters $\beta$ 
\cite{Gurarie2, GurarieLudwig, MathieuRidout, Ridout2, Dubail_beta, Vasseur_beta}) that multiply the logarithms, can only be given in $d=2$. The reason that our results differ from
(\ref{2ptPolyakov})--(\ref{3ptPolyakov}) can be traced back to 
one key assumption made by Polyakov, which is that the operators are quasi-primary.
In our LCFT context this assumption would be incorrect, since two (or more) operators having the same scaling dimension may be linked up in an indecomposable Jordan cell structure, that is, only one of the fields in each Jordan cell is an eigenstate of the dilatation operator.

The paper is laid out as follows. In section 2, we classify the scalar operator content of the Potts model using the representation theory of the symmetric group. We then give in section 3 some exact expressions -- valid on an arbitrary graph --for the correlation functions of these operators in terms of probabilities. The consequences of these lattice results on the scaling limit are then discussed in section 4. Sections 5, 6, and 7 contain a detailed analysis of the limits $Q \to 0,1,2$, respectively, along with physical applications to spanning trees and forests, resistor networks, bond percolation, and the Ising model. In section 8, we show how our analysis can be used to deduce (some parts of) the general logarithmic structure of the Potts field theory in arbitrary dimension. Finally, section 9 contains a short discussion of our results and of the perspectives for further studies.

\section{Classification of the operators in the Potts model}

The $Q$-state Potts model plays a central role in statistical
physics. Its partition function and operator content can be
represented in a variety of ways, in terms of $Q$-component spins,
clusters, loops, vertex models, and heights. Some of these
representations are particular for $d=2$ dimensions, while other are
valid for any $d$. Note also that although $Q$ is initially assumed to
be a positive integer, many of the representations make sense for
arbitrary real $Q$ by suitable analytic continuations.

Below we review the necessary prerequisites on the Potts model, with
an emphasis on the role of the symmetric group $S_Q$. We then proceed
to classify its (scalar) operator content in arbitrary dimension, in terms of
representations of $S_Q$. These representation theoretical
preliminaries will turn out instrumental for establishing the
logarithmic nature of certain geometrically defined correlation
functions. Also, the possibility of making analytic continuations in
$Q$ is crucial for certain limiting procedures that are at the heart
of our subsequent treatment of the continuum limit. 

\subsection{Potts model and symmetric group $S_Q$}

The $Q$-state Potts model on a graph $G=(V,E)$ is defined through
$Q$-component spins $\sigma_i = 1,2,\ldots,Q$ that live on the
vertices $i \in V$ and interact along the edges $(ij) \in E$ via an
interaction energy $-K \delta_{\sigma_i,\sigma_j}$ proportional to the
Kronecker symbol ($\delta_{x,y} = 1$ if $x=y$, and 0 otherwise).  Its
partition function thus reads
\be
 Z = \sum_{\sigma} \prod_{(ij) \in E} {\rm e}^{K \delta_{\sigma_i,\sigma_j}} \,,
 \label{Zspin}
\ee
where the sum is over all spins $\sigma = \{ \sigma_i | i \in V \}$.
In most applications $G$ is a regular lattice in $d$ dimensions.

Since $\delta_{x,y}$ can only take two values, the identity ${\rm
  e}^{K \delta_{\sigma_i,\sigma_j}} = 1 + v
\delta_{\sigma_i,\sigma_j}$ holds with $v = {\rm e}^K - 1$. Expanding
out the product $\prod_{(ij) \in E}$ one gets a sum over subsets $A
\subseteq E$ of edges for which the term $v
\delta_{\sigma_i,\sigma_j}$ is taken. Each connected component
(including isolated vertices) in the spanning subgraph $(V,A)$ is
called a Fortuin-Kasteleyn (FK) cluster. The factors of
$\delta_{\sigma_i,\sigma_j}$ entail that the spin is constant on each
FK cluster, so performing the sum $\sum_{\sigma}$ results in \cite{FK}
\be
 Z = \sum_{A \subseteq E} Q^{k(A)} v^{|A|} \,,
 \label{ZFK}
\ee
where $k(A)$ is the number of connected components in the spanning
subgraph $(V,A)$ with $|A|$ edges.

\begin{figure}
\centering
\includegraphics[width=7cm]{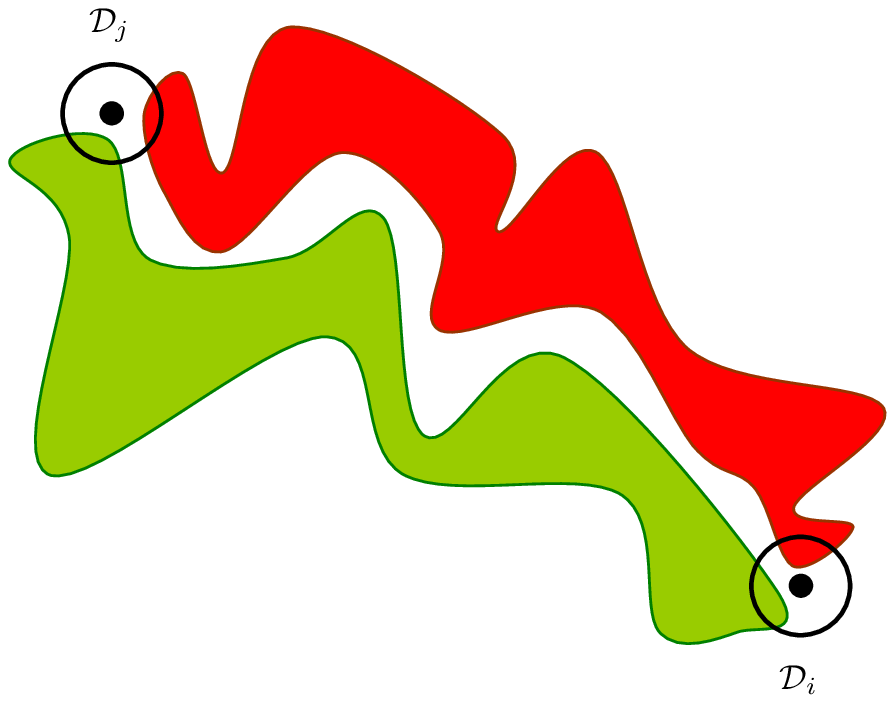}\hspace*{0.5cm}
\includegraphics[width=7cm]{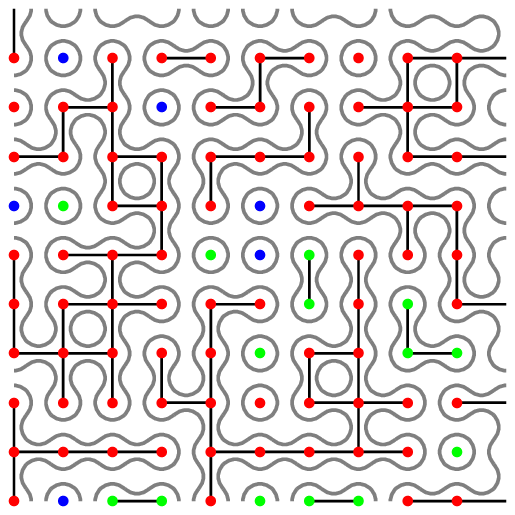}
\caption{Left: clusters connectivity corresponding to the 2-cluster operator. Right: Spin configuration of the two-dimensional three-state Potts model ($Q=3$), and corresponding FK clusters and dual loops.}\label{FigFK}
\end{figure}

The FK representation (\ref{ZFK}) of the partition function is valid
for any graph, thus in particular for lattices in any dimension
$d$ (see Fig.~\ref{FigFK} in $d=2$). It has the advantage over (\ref{Zspin}) that $Q$ appears as a
formal parameter, making it possible to approach physical (i.e.,
integer) values via a limiting procedure. However, the form
(\ref{ZFK}) obscures the $S_Q$ symmetry of permuting the spin labels
that was initially manifest in (\ref{Zspin}). This symmetry content
needs to be accounted for when discussing correlation functions (see
below).

We stress that while all spins on a given FK cluster are identical,
spins on different FK clusters have been independently summed over to
obtain (\ref{ZFK}). In particular, two distinct FK clusters may or may
not carry the same spin value, even when they are adjacent in $G$. It
is possible to define ``spin clusters'' as connected regions in $G$
with constant spin, even when $Q$ is not integer \cite{DJS,VJ}, but in
this paper we shall exclusively consider the more well-known FK
clusters.

We also remark that when $d=2$ many other representations of $Z$ are
possible. Among those, the {\em loop representation} consists in
trading the FK clusters for their surrounding (inner and outer)
contours on the medial graph ${\cal M}(G)$. Using topological
identities this results in \cite{BKW}
\be
 Z = Q^{|V|/2} \sum_{A \subseteq E} Q^{\ell(A)/2}
 \left( \frac{v}{\sqrt{Q}} \right)^{|A|} \,,
 \label{eqZloop}
\ee
where $\ell(A)$ is the number of circuits (or closed loops) formed by
the cluster contours.

A very useful correlation function---that we shall study extensively
below---is given by the probability that $N$ distinct FK clusters
extend from a small neighborhood ${\cal D}_i$, containing $N$ vertices
and centered at $r_i$, to another similar neighborhood ${\cal D}_j$,
centered at a distant point $r_j$. This is again defined in any
dimension $d$. For $d=2$ this means that $2N$ loop strands come close
in ${\cal D}_i$ and again in ${\cal D}_j$ (see Fig.~\ref{FigFK}). This situation is known in
the literature as a $N$-cluster (or $2N$-watermelon in $d=2$ dimensions) configuration. The cluster
configuration can be viewed as a topological defect, implying in particular that 
the $N$ clusters ``inserted'' at ${\cal D}_i$ cannot contract among themselves,
but must propagate until they are ``taken out'' at ${\cal D}_j$.

By extension, the operator that ensures that $N$ points in a small
neighborhood belong to distinct propagating FK clusters shall henceforth be called
a {\em cluster operator}, also when $d>2$. Correlation functions of
cluster operators will play a paramount role in the remainder of
this paper.

When the Potts model stands at a second order phase transition the
long-distance behavior of correlation functions will generically be
power laws of the separations $r_{ij} = |r_i - r_j|$ between the small
neighborhoods. However, when $Q$ takes particular values---and when
considering appropriate correlation functions---these power laws may
be multiplied by logarithms. When $d=2$ the critical exponents
entering the power laws can be found exactly from conformal field
theory (CFT), and in some cases (such as $Q=2$~\cite{Smirnov}) even rigorously by
proving that the Potts model converges to a stochastic process known
as Schramm-Loewner Evolution (SLE) in the continuum limit~\cite{SLE1,SLE2}. 
More recently, the logarithmic part of the $d=2$
correlations has been fixed from considerations on logarithmic
conformal field theory (LCFT)~\cite{LCFT1,LCFT2,LCFT3}.

The main purpose of this paper is to make precise the logarithmic
nature of the correlation functions of cluster operators in any
dimension $d > 2$. We shall see that the necessary\footnote{Note that strictly speaking, these conditions are not sufficient in general (see section~\ref{secQ0d3}).}algebraic conditions for producing logarithms (i.e., a particular mixing of
operators, to be made precise below) is operative whenever $Q \to Q_0$,
with $Q_0$ a non-negative integer. Although we expect effects of this
mixing for any integer $Q_0 \ge 0$, we shall eventually constrain the
detailed discussion to the cases $Q_0 = 0$ (spanning trees and
forests), $Q_0 = 1$ (percolation), and $Q_0 = 2$ (Ising model), since
in these cases the Potts model is known \cite{Amit,Forests3D} to stand
at a non-trivial (i.e., non-mean field) second order phase transition
for any dimension $2 \le d < d_{\rm uc}$, where $d_{\rm uc} = 6$
denotes the upper critical dimension for $Q<2$ and $d_{\rm uc}=4$ for Ising.
Note that the $Q=3$-state model is 
known to be non-critical already in $d=3$~\cite{Knak,Gavai}. 

In $d=2$ the corresponding critical temperatures $v_{\rm c}$ are known
exactly for the square lattice ($v_{\rm c} = \sqrt{Q}$), the
triangular lattice ($v_{\rm c} > 0$ solution of $v^3 + 3 v^2 = Q$),
and the hexagonal lattice (the dual of the former), and with
considerable precision for a variety of other lattices
\cite{JS2012}. In higher dimension $v_{\rm c}$ is known approximately
from a variety of numerical and perturbative techniques~\cite{Ising3D, Perco3D, Forests3D, IsingAus}.

\subsection{Representation theory}
\label{sec:repr_th}

As a first step in constructing the $N$-cluster (watermelon) operators
in arbitrary dimension, we need some representation theoretical
preliminaries. The intuitive idea is that we must construct a certain
operator acting on $N$ spins (that will eventually be taken to be
adjacent, or belong to a small neighborhood) and satisfying some
symmetry requirement. If we require the $N$ spins to take different
values they will obviously belong to distinct FK clusters, but this is
not enough, since we must also ensure that the $N$ FK clusters inserted
by the cluster operator persist until they are taken out by another
cluster operator (and not just ``end in the middle of nowhere'').

To make this idea precise, and to attain our objectives, we now set
out to classify $N$-spin operators from the point of view of
irreducible representations (irreps) of the symmetric group $S_Q$.

Let $L^{(1)}_{Q}$ be the span of all $Q$-component vectors
$\mathcal{O}(\sigma)$, where $\sigma=1,\dots,Q$. This space obviously
has dimension $Q$.  Letting the symmetric group $S_Q$ acting on the
index $\sigma$, this defines a reducible representation of this group.
It is indeed straightforward to form the invariant quantity
$\sum_{\sigma=1}^{Q} \mathcal{O}(\sigma)$ which transforms as the
trivial, one-dimensional irrep $\left[ Q\right]$.  The decomposition
with respect to $S_Q$ thus yields
\begin{equation}
L^{(1)}_{Q} = \left[ Q\right] \oplus \left[ Q-1,1 \right],
\end{equation}
where $\left[ Q-1,1 \right]$ is the space of vectors
$\mathcal{O}(\sigma)$ that satisfy $\sum_{\sigma}
\mathcal{O}(\sigma)=0$, with dimension $Q-1$.

Here, and in the sequel, an irrep $\lambda$ of $S_Q$ is denoted by its
Young diagram $[\lambda_1,\lambda_2,\ldots,\lambda_k]$, where $\lambda_i$
is the length of row $i$. It is a remarkable fact that we shall only
need representations with at most $k=2$ rows; the same property is well
known to hold in $d=2$ dimensions \cite{Moore}.

Let us now consider a slightly more involved example. Let
$L^{(2)}_{Q}$ be the span of all symmetric $Q \times Q$ matrices
$\mathcal{O}(\sigma_1,\sigma_2)$ with zero elements on the
diagonal. We have $\mathrm{dim} \ L^{(2)}_{Q} =
\frac{Q(Q-1)}{2}$. Once again, acting with $S_Q$ on the indices
$\sigma_1$ and $\sigma_2$, this defines a representation which turns
out to be reducible. Indeed, the subspace of matrices that satisfy
$\sum_{\sigma_1=1}^{Q} \mathcal{O}(\sigma_1,\sigma_2) =0$ provides an
irreducible representation $\left[ Q-2,2 \right]$ of dimension
$\frac{Q(Q-1)}{2}-Q=\frac{Q(Q-3)}{2}$.  The quotient of $L^{(2)}_{Q}$
by this invariant subspace is a representation of dimension $Q$,
isomorphic to $L^{(1)}_{Q}$, which is spanned by the vector
$\tilde{\mathcal{O}}(\sigma_2) = \sum_{\sigma_1}
\mathcal{O}(\sigma_1,\sigma_2)$.  We therefore write
\begin{equation}
L^{(2)}_{Q} = \underbrace{\left[ Q\right] \oplus \left[ Q-1,1 \right]}_{ \tilde{\mathcal{O}}(\sigma_2) = \sum_{\sigma_1} \mathcal{O}(\sigma_1,\sigma_2)}  \oplus \underbrace{\left[ Q-2,2 \right] }_{{\sum_{\sigma_1} \mathcal{O}(\sigma_1,\sigma_2) =0}}.
\end{equation}  

The generalization of this decomposition is straightforward. For
example, the space $L^{(3)}_{Q}$ of $Q\times Q \times Q$ symmetric
tensors $\mathcal{O}(\sigma_1,\sigma_2,\sigma_3)$ which vanish when
two indices coincide is decomposed as $ L^{(2)}_{Q} \oplus \left[
  Q-3,3 \right] $, where the subspace $\left[ Q-3,3 \right]$
corresponds to the tensors that satisfy the $Q(Q-1)/2$ constraints
$\sum_{\sigma_1} \mathcal{O}(\sigma_1,\sigma_2,\sigma_3)=0$. The
remaining space $L^{(2)}_{Q}$ then corresponds to the decomposition of
$\tilde{\mathcal{O}}(\sigma_2,\sigma_3) = \sum_{\sigma_1}
\mathcal{O}(\sigma_1,\sigma_2,\sigma_3)$.  In general, we will denote
by $L^{(N)}_{Q}$ the space of $Q\times Q\times \dots \times Q$
symmetric tensors of rank $N$ that vanish whenever two indices
coincide.  The dimension of $L^{(N)}_{Q}$ is $\frac{Q(Q-1) \dots
  (Q-N+1)}{N!}$.  By induction we then have the decomposition
\begin{equation}
L^{(N)}_{Q} = \underbrace{\left[ Q\right] \oplus \left[ Q-1,1 \right] \oplus \dots \oplus \left[ Q-N+1,N-1 \right]}_{ \tilde{\mathcal{O}}(\sigma_2,\dots,\sigma_N) = \sum_{\sigma_1} \mathcal{O}(\sigma_1,\sigma_2,\dots,\sigma_N)}  \oplus \underbrace{\left[ Q-N,N \right] }_{{\sum_{\sigma_1} \mathcal{O}(\sigma_1,\sigma_2,\dots,\sigma_N) =0}}.
\end{equation}

\subsection{$N$-cluster operators $t^{(k,N)}_{a_1,\dots,a_k}$}
\label{sec:watermelon}

We consider a symmetric operator $\mathcal{O}(\sigma_1,\dots,\sigma_N)$, 
defined on $N$ distinct Potts spins, and we impose that it vanishes 
if any of the $N$ spins coincide. We would like to understand how to decompose
this operator in terms of irreps of the symmetric group.
According to the results of the previous section, this means that
we want to construct each of the representations in the decomposition
\begin{equation}
 [Q] \oplus [Q-1,1] \oplus [Q-2,2] \oplus \ldots \oplus [Q-N,N].
 \label{decomp}
\end{equation}
By the hook formula the dimensions of the representations are
\begin{equation}
 d_k \equiv {\rm dim}([Q-k,k]) = \frac{Q!}{(Q-k+1)!} \,
 \frac{Q-2k+1}{k!} \,,
\end{equation}
and in terms of dimensions the decomposition (\ref{decomp}) reads
\begin{equation}
 D_N \equiv \sum_{k=0}^N d_k = (1) + (Q-1) + \left( \frac{Q(Q-3)}{2} \right)
 + \ldots + \left( \frac{Q!}{(Q-N+1)!} \,
 \frac{Q-2N+1}{N!} \right) = \frac{Q!}{(Q-N)! \, N!} \,.
 \label{DN}
\end{equation}
This is indeed the number of symmetric tensors that vanish if any
two spins coincide.

\subsubsection{Constructing the invariant tensors}

To construct an explicit basis for these representations, we proceed
as follows.  Let us consider first the case $N=1$. The invariant $[Q]$
is just a constant $t^{(0,1)} \equiv 1 = \sum_{a} \delta_{a,\sigma_1}$
in that case. Meanwhile, the $Q-1$ generators of the irrep $[Q-1,1]$
read
\begin{equation}
 t_{a}^{(1,1)}(\sigma_1) = \delta_{a,\sigma_1} - \frac{1}{Q} t^{(0,1)} \,.
 \label{t11}
\end{equation}
These operators satisfy $\sum_a t_{a}^{(1,1)}=0$ so we indeed have
only $Q-1$ of them, and note that we also have $\sum_{\sigma_1}
t_{a}^{(1,1)}(\sigma_1)=0$ which is expected by the definition of the
representation $[Q-1,1]$.

We next consider the case $N=2$. The invariant $[Q]$ is nothing but
$t^{(0,2)} \equiv \delta_{\sigma_1 \neq \sigma_2} =
1-\delta_{\sigma_1,\sigma_2}$. Note that since we already decided that
our operators vanish whenever two spins coincide, we can set
$t^{(0,2)}=1$ in that space. The representation $[Q-1,1]$ is trivially
obtained from the case $N=1$ as
\begin{equation}
 t_{a}^{(1,2)}(\sigma_1,\sigma_2) = t_{a}^{(1,1)}(\sigma_1)+ t_{a}^{(1,1)}(\sigma_2) = \delta_{\sigma_1,a}+ \delta_{\sigma_2,a} - \frac{2}{Q}.
 \label{ta12}
\end{equation}
The only new non-trivial case is the basis of $[Q-2,2]$. We are looking for a basis of $Q(Q-3)/2$ operators $t^{(2,2)}(\sigma_1,\sigma_2)$
that satisfy
\be
 \sum_{\sigma_1} t_{ab}^{(2,2)}(\sigma_1,\sigma_2) = 0 \,,
 \label{t22constraint}
\ee
as explained in section~\ref{sec:repr_th}. We label them using two
symmetric indices $a,b$ that run from $1$ to $Q$. It is clear that
$t^{(2,2)}(\sigma_1,\sigma_2)$ must contain the two terms
$\delta_{\sigma_1,a}\delta_{\sigma_2,b} +
\delta_{\sigma_2,a}\delta_{\sigma_1,b}$. However, just as in
(\ref{ta12}) we need to subtract multiples of the lower-order tensors
in order the fulfill the constraint (\ref{t22constraint}). Solving the
resulting linear system we easily find
\begin{equation}
 t_{ab}^{(2,2)}(\sigma_1,\sigma_2) = \delta_{\sigma_1,a}\delta_{\sigma_2,b} + \delta_{\sigma_2,a}\delta_{\sigma_1,b} - \frac{1}{Q-2} \left( t_{a}^{(1,2)}(\sigma_1,\sigma_2)+t_{b}^{(1,2)}(\sigma_1,\sigma_2) \right) - \frac{2}{Q(Q-1)} t^{(0,2)} ,
 \label{tab22}
\end{equation}
for $a \neq b$; when $a=b$ we have $t_{ab}^{(2,2)}(\sigma_i,\sigma_j)
= 0$ by definition. One can check that the constraint holds also for
the tensor indices, namely $\sum_a t_{ab}^{(2,2)}=0$, so that there
are $Q(Q-1)/2-Q=Q(Q-3)/2$ generators indeed.

\subsubsection{General procedure}

The general pattern should already be clear at this point. We proceed
inductively. On $N$ spins, one can use the results for fewer spins to
construct the tensors $t^{(k,N)}_{a_1,\dots,a_k}$, with $k=1,\dots,
N-1$. The non-trivial step consists in finding the basis
$t^{(N,N)}_{a_1,\dots,a_N}$ of the last representation $[Q-N,N]$. It
is obtained from the properly symmetrized combination of Kronecker
deltas by imposing the constraint
\be
 \sum_{\sigma_1} t^{(N,N)}_{a_1,\dots,a_N}(\sigma_1,\dots,\sigma_N) = 0 \,.
 \label{tNNconstraint}
\ee
Let us spell out this procedure in full details. The Ansatz for
$t^{(N,N)}$ involves the subtraction of $N$ terms of the type
$t^{(k,N)}$ with $k=0,1,\ldots,N-1$. Each of the subtracted terms
depends only on a subset of $k$ out of the $N$ available tensor
indices, and must therefore by symmetrized over the ${N \choose k}$
possible ways of forming these subsets. Then, for any choice of
distinct tensor indices $\{a_1,\ldots,a_N\}$, the constraint
(\ref{tNNconstraint}) results in a distinct equation according to how
many of the free spin indices $\{\sigma_2,\ldots,\sigma_N\}$ coincide
with one of the $a_i$. This number is $l = 0,1,\ldots,N-1$.
Thus, we recover precisely $N$ independent
linear equations, sufficient to solve for the coefficients in front of
each of the $N$ subtracted tensor terms. Once the result for
$t^{(N,N)}$ has been obtained, one can then verify that the
tensor-index analogue of the (\ref{tNNconstraint}), namely
\be
 \sum_{a_1} t^{(N,N)}_{a_1,\dots,a_N}(\sigma_1,\dots,\sigma_N) = 0 \,,
\ee
holds as well.

\subsubsection{General results for the invariant tensors}
\label{sec:tensor_results}

In order to state the results for generic $N$, we switch to a less
cumbersome notation so that no indices appear.  For example for $N=1$,
we will denote the result (\ref{t11}) as
\begin{equation}
 t^{(1,1)} = (1 \delta) - \frac{1}{Q} \left(1 t^{(0,1)} \right)
\end{equation}
while for $N=2$, we denote the results (\ref{ta12}) and (\ref{tab22}) as
\begin{eqnarray}
 t^{(1,2)} &=& (2 \delta) - \frac{2}{Q} \left(1 t^{(0,2)} \right) \,, \nonumber \\
 t^{(2,2)} &=& (2 \delta) - \frac{1}{Q-2} \left( 2 t^{(1,2)} \right) - \frac{2}{Q(Q-1)} \left(1 t^{(0,2)} \right) \,.
\end{eqnarray}
The integer appearing in front of $\delta$ and $t_i$, inside a pair of
parentheses, indicates how many terms participate in the suitably
symmetrized object. The coefficients appearing outside such
parentheses are those worked out by the procedure explained
above. This succinct notation keeps explicit the pole structure in $Q$
which will be the most important point in the remainder of this paper.

We have continued this construction explicitly for higher $N$. For
$N=3$ spins we find
\begin{eqnarray}
 t^{(1,3)} &=& (3 \delta) - \frac{3}{Q} \left(1 t^{(0,3)} \right) \,, \nonumber \\
 t^{(2,3)} &=& (6 \delta) - \frac{2}{Q-2} \left( 2 t^{(1,3)} \right) - \frac{6}{Q(Q-1)} \left(1 t^{(0,3)} \right) \,, \\
 t^{(3,3)} &=& (6 \delta) - \frac{1}{Q-4} \left( 3 t^{(2,3)} \right) -
   \frac{2}{(Q-2)(Q-3)} \left( 3 t^{(1,3)} \right) - \frac{6}{Q(Q-1)(Q-2)} \left(1 t^{(0,3)} \right) \,. \nonumber
\end{eqnarray}
The conjectured general result (which we have confirmed by working out the $N=4$
case in details) takes the form:
\begin{equation}
 t^{(k,N)} = \left( \alpha_k \delta \right) -
  \sum_{i=0}^{k-1} \gamma_{k,i} \left( \beta_{k,i} t^{(i,N)} \right)
 \label{general_tensor}
\end{equation}
with the coefficients
\begin{eqnarray}
 \alpha_k &=& \frac{N!}{(N-k)!} \,, \nonumber \\
 \beta_{k,i} &=& \frac{k!}{(k-i)! \, i!} \,, \\
 \gamma_{k,i} &=& \frac{(N-i)!}{(N-k)!} \, \frac{(Q-i-k)!}{(Q-2i)!} \,. \nonumber
\end{eqnarray}

We now claim that for a given number of spins $N$, the most symmetric
tensor $t^{(N,N)}$ is the $N$-cluster operator in arbitrary
dimension $d$. This statement will be corroborated in
section~\ref{sec:discrete} where we show that the corresponding
two-point functions are proportional to the probability that $N$
distinct FK clusters connect each of the two groups of $N$
points.   We defer the precise
interpretation of the tensors of lower rank, $t^{(k,N)}$ with $k < N$,
to section~\ref{secSubleading}.

\subsubsection{Case of unconstrained tensors}

The general results of section~\ref{sec:tensor_results} pertain to
tensors which are constrained to be symmetric in all spin variables
and to vanish if any of these coincide. While these constrained
tensors are the only ones to be used in the remainder of this paper, it is interesting to consider also the class of
tensors without this constraint.

Let us first remark that the number of ways that $N$ spins can take
exactly $p \le N$ distinct values is equal to the number of ways that
an $N$-element set can be partitioned into $p$ non-empty parts.  This
latter number, denoted $S_2(N,p)$, is called the Stirling number of the second kind 
and reads explicitly
\be
 S_2(N,p) = \frac{1}{p!} \sum_{k=0}^p (-1)^k {p \choose k} (p-k)^N \,.
\ee
Next, the number of
tensors acting on $p$ such groups of spins, and symmetric under the
interchange of any two groups, is obviously $D_p$ given by
(\ref{DN}). Without the symmetrization we would have $p! D_p$
tensors. The number of tensors with any number of groups of coincident
spins is therefore
\be
 \sum_{p=1}^N p! \, D_p S_{N,p} = Q^N \,.
\ee
There are thus exactly $Q^N$ unconstrained tensors, as expected.

\section{Correlation functions: discrete results}
\label{sec:discrete}

We next show that one can obtain useful structural results on
correlation functions of the tensors $t^{(k,N)}$ constructed in
section~\ref{sec:watermelon}, by combining the representation
theoretical tools of section~\ref{sec:repr_th} with elementary
combinatorial considerations. These results account in particular for
the dependence of correlation functions on the tensorial
indices. Moreover, the correlation functions of cluster operators
will be related to linear combinations of the probabilities that the
spins acted on by one operator are connected by FK clusters to spins
acted on by other operators in various ways. This geometrical content
is essential for unraveling the physical interpretation of the
correlation functions.

The structural results found in this section involve various
coefficients that have poles when $Q$ tends to a non-negative integer. The
cancellation of these singularities is at the heart of the mechanism
that will eventually produce the logarithmic behavior of correlation
functions in the continuum limit.

Throughout this section we only apply combinatorial considerations to
the finite number of spins that enter explicitly in the cluster
operators. The results are therefore completely general and do not, in
particular, depend on the graph (or $d$-dimensional lattice) on which
the Potts model is defined.

\subsection{Two-point functions: $N=1$ spin}

We recall from section~\ref{sec:watermelon} that the two operators
acting on one spin read, in our notation, $t^{(0,1)}(\sigma_1) = 1$
and $t^{(1,1)}_a(\sigma_1) = \delta_{a,\sigma_1}-\frac{1}{Q}$. The
two-point functions of these operators are the following:
\begin{align}
\left\langle t^{(0,1)} (r_1) t^{(0,1)} (r_2) \right\rangle &= 1 \,, \label{t01t01} \\
\left\langle t_{a}^{(1,1)} (r_1) t_{b}^{(1,1)} (r_2) \right\rangle &= \frac{1}{Q} \left(\delta_{a,b} - \frac{1}{Q} \right) \cc \,. \label{t11t11}
\end{align}
While the former result is of course trivial, we wish to spend a moment
discussing the latter result in order to carefully fix some ideas and
notations to be used throughout this section.

First, we imagine that the two groups of $N=1$ spins are situated at (or
later, when $N>1$, in small neighborhoods around) the points $r_1$ and
$r_2$ respectively. Obviously we cannot specify how the correlation
function depends on these coordinates, since we have not yet assumed
anything about the lattice on which the Potts model is defined, nor
whether the coordinates are widely separated. Such input is deferred
to section~\ref{sec:CFT}, where we shall start exploiting the consequences of scale
and conformal invariance. However, we can still denote by $\notccsmall$ the
probability that the two spins belong to two different FK clusters, and by
$\ccsmall$ the probability that they belong to the same FK
cluster.  In the former case, the two spins are summed over
independently, and the coefficient of $\notccsmall$ is
\be
 \frac{1}{Q^2} \sum_{\sigma_1,\sigma_2}
 \left( \delta_{a,\sigma_1} - \frac{1}{Q} \right)
 \left( \delta_{b,\sigma_2} - \frac{1}{Q} \right) = 0 \,.
 \label{t11t11A}
\ee
In the latter case, the two spins are constrained to take the same
value, and the coefficient of $\ccsmall$ is therefore
\be
 \frac{1}{Q} \sum_{\sigma_1}
 \left( \delta_{a,\sigma_1} - \frac{1}{Q} \right)
 \left( \delta_{b,\sigma_1} - \frac{1}{Q} \right) =
 \frac{1}{Q} \left( \delta_{a,b} - \frac{1}{Q} \right) \,.
 \label{t11t11B}
\ee
Combining (\ref{t11t11A})--(\ref{t11t11B}) we arrive at (\ref{t11t11}).

We also note that the mixed correlation function is identically zero:
\be
 \left\langle t^{(0,1)}(r_1) t_a^{(1,1)}(r_2) \right\rangle = 0 \,.
\ee
This is a general feature that will carry over to higher $N$ for symmetry reasons.

\subsection{Two-point functions: $N=2$ spins}

We now move to the slightly more involved case of $N=2$ spins.  Our
results read as follows:
\begin{align}
\left\langle t^{(0,2)} (r_1) t^{(0,2)} (r_2) \right\rangle &= \left( \frac{Q-1}{Q} \right)^2 \left(\Pc+\Pb  \right) + \frac{Q-1}{Q} \Pa, \label{t02t02} \\
\left\langle t_{a}^{(1,2)} (r_1) t_{b}^{(1,2)} (r_2) \right\rangle &= \frac{Q-2}{Q^2} \left(\delta_{a,b} - \frac{1}{Q} \right)\left( \frac{Q-2}{Q}\Pb + 2 \Pa \right), \label{t12t12} \\
\left\langle t_{ab}^{(2,2)} (r_1) t_{cd}^{(2,2)} (r_2) \right\rangle &= \frac{2}{Q^2} \left( \delta_{ac} \delta_{bd} + \delta_{ad} \delta_{bc} - \frac{1}{Q-2} (\delta_{ac} +\delta_{bd} + \delta_{ad}+ \delta_{bc})+ \frac{2}{(Q-2)(Q-1)} \right) \Pa. \label{t22t22}
\end{align}
In the corresponding diagrams, the spins corresponding to the leftmost
operator (and that we imagine situated in a neighborhood around $r_1$)
are shown on the bottom, and those corresponding to the rightmost
operator are depicted on the top. We denote by $\Pasmall$ (resp.\ $\Pbsmall$,
or $\Pcsmall$) the probability that there are zero (resp.\ one, or two) FK clusters
connecting an $r_1$ point to an $r_2$ point. Notice that the different
points belonging to the same operator cannot be connected among
themselves, because of the constraint that the cluster operators
vanish in the case of coinciding spins. We also stress that $\Pbsmall$
is the probability that any one of the two $r_1$ points
is in the same FK cluster as any one of the two $r_2$ points, so even
though the connected pair of points is shown on the left, the diagram
actually stands for a sum of four distinct contributions. This is consistent
with the fact that the cluster operators are symmetric in their
spin indices.

To fix the coefficients appearing in front of the three probabilities in
(\ref{t02t02}) one simply needs to average the product of
$t^{(0,2)}(r_1) = 1 - \delta_{\sigma_1,\sigma_2}$ and $t^{(0,2)}(r_2)
= 1 - \delta_{\sigma_3,\sigma_4}$ over the spins
$\sigma_1,\sigma_2,\sigma_3,\sigma_4$, upon inserting an extra factor
of $1$ in the case of $\Pcsmall$, a factor $\delta_{\sigma_1,\sigma_3}$
in the case of $\Pbsmall$, and a factor $\delta_{\sigma_1,\sigma_3}
\delta_{\sigma_2,\sigma_4}$ in the case of $\Pasmall$. Doing this leads to
the result shown in (\ref{t02t02}).

To establish (\ref{t12t12})--(\ref{t22t22}) one further needs to take account
of the tensor indices. It is useful to write first an Ansatz for the possible
dependence on the tensor indices. In the case of (\ref{t12t12}) this is obvious
provided by $c_{1,1} \delta_{a,b} - c_{0,1}$, where $c_{1,1}$ and $c_{0,1}$
are some constants. We then apply the calculational scheme
just explained for the $t^{(0,2)}$ correlator to the case at hand where
$t_a^{(1,2)}$ is given by (\ref{ta12}); the two cases $a=b$ and $a \neq b$
must be examined in turn to fix both constants $c_{1,1}$ and $c_{0,1}$.
In the case of (\ref{t22t22}) the Ansatz for the dependence on the tensor
indices should obviously take into account that $t_{ab}^{(2,2)}$ is zero
when $a=b$. Therefore we have an Ansatz of the type
\be
 c_{2,2} (\delta_{ac} \delta_{bd} + {\rm perm}) +
 c_{1,2} (\delta_{ac} + {\rm perm}) + c_{0,2} \,,
\ee
where ${\rm perm}$ denotes all internal permutations among the indices
in the $r_1$ operator, and among those in the $r_2$ operator. To fix the three
constants, the calculation must be done in the cases where the values of
the indices $a,b$ coincide with zero, one or two of the indices $c,d$.
Obviously the structure of Kronecker deltas acting on the tensor indices
is very reminiscent of the one appearing in the spin variable probabilities
$\Pasmall$, $\Pbsmall$ and $\Pcsmall$.

The results (\ref{t02t02})--(\ref{t22t22}) display a remarkable feature that
will carry over to higher $N$ as well. Namely, the two-point function of
the cluster operator $t^{(p,N)}$ couples only to probabilities that
there are {\em at least $p$ distinct FK clusters} connecting the set of points
in the first and the second operator. In particular,
$\langle t^{(N,N)}(r_1) t^{(N,N)}(r_2) \rangle$
is proportional to the probability of having $N$ propagating FK clusters.
This establishes our claim that $t^{(N,N)}$ is the $N$-cluster 
operator. The precise interpretation of $t^{(p,N)}$ for $p<N$ is more tricky
and will be deferred to section~\ref{secSubleading}; suffice it here to say that loosely
speaking this operator inserts ``at least'' $p$ propagating FK clusters.

We have checked that just like in the $N=1$ case all mixed correlation functions vanish:
\be
 \left\langle t^{(0,2)}(r_1) t_a^{(1,2)}(r_2) \right\rangle =
 \left\langle t^{(0,2)}(r_1) t_{ab}^{(2,2)}(r_2) \right\rangle =
 \left\langle t_a^{(1,2)}(r_1) t_{bc}^{(2,2)}(r_2) \right\rangle = 0 \,.
 \label{mixedcorr2}
\ee
This result could in fact be established without resorting to explicit calculations, since
$t^{(0,2)}$, $t^{(1,2)}$ and $t^{(2,2)}$ have been constructed as different irreps of $S_Q$;
the vanishing of mixed correlations then follows from representation theoretical reasons.
But (\ref{mixedcorr2}) is of course also consistent with the features mentioned in the preceding
paragraph. Namely the vanishing
correlator of the product between $t^{(0,2)}$ and either $t^{(1,2)}$ or $t^{(2,2)}$ is
due to the fact that $t^{(0,2)}$ cannot ``take out'' the FK clusters ``inserted'' by either
of the two latter operators. However the vanishing of $\langle t^{(1,2)} t^{(2,2)} \rangle $
cannot be explained from this simple reasoning.

This type of reasoning (or explicit calculations) also imply that the one-point functions
of $t^{(p,N)}$ vanish for $p>0$; in particular
$\langle  t_a^{(1,2)}(r) \rangle = \langle  t_{ab}^{(2,2)}(r) \rangle = 0$.
However, $\langle t^{(0,2)} \rangle = \langle 1-\delta_{\sigma_1,\sigma_2} \rangle$
is non-zero and is proportional to the probability $\Pdiff$ that the spins $\sigma_1,\sigma_2$
belong to two different FK clusters. It is convenient to define an operator with this mean value
subtracted:
\begin{equation}
\epsilon (r) \equiv  t^{(0,2)}(r) - \frac{Q-1}{Q} \Pdiff \,,
\label{energydef}
\end{equation}
so that now $\langle \epsilon(r) \rangle = 0$. We shall see in section~\ref{sec:CFT} that in
the continuum limit $\epsilon (r)$ is proportional to the energy operator, as indicated
by the chosen notation. Using again the same methods as above, we then find that
\begin{equation}
\langle \epsilon (r_1) \epsilon (r_2)\rangle = \frac{Q-1}{Q} \left(  \Pa + \left( 1-\frac{1}{Q} \right) \left(\Pc+\Pb -  \Pdiff^2 \right)\right) \,.
\end{equation}

\subsection{Two-point functions: $N=3$ spins}

The general method and the structure of the results is now clear. For
$N=3$ we find the following two-point functions:
\begin{align}
\left\langle t^{(0,3)} (r_1) t^{(0,3)} (r_2) \right\rangle &=  \frac{(Q-2)(Q-1)}{Q^2} \left( \frac{(Q-2)(Q-1)}{Q^2} \left(\PPPd+\PPPc \right)+  \frac{Q-2}{Q}\PPPb+\PPPa \right), \label{t03t03} \\
\left\langle t_{a}^{(1,3)} (r_1) t_{b}^{(1,3)} (r_2) \right\rangle &= \frac{(Q-3)(Q-2)}{Q^3} \left(\delta_{a,b} - \frac{1}{Q} \right)\left(\frac{(Q-3)(Q-2)}{Q^2}\PPPc+ \frac{2 (Q-3)}{Q}\PPPb + 3 \PPPa \right), \\
\left\langle t_{ab}^{(2,3)} (r_1) t_{cd}^{(2,3)} (r_2) \right\rangle &= \frac{2(Q-4)}{Q^3} \left( \delta_{ac} \delta_{bd} + \delta_{ad} \delta_{bc} - \frac{1}{Q-2} (\delta_{ac} +\delta_{bd} + \delta_{ad}+ \delta_{bc})+ \frac{2}{(Q-2)(Q-1)} \right) \notag \\
& \times \left( \frac{Q-4}{Q}\PPPb+3 \PPPa \right), \\
\left\langle t_{abc}^{(3,3)} (r_1) t_{def}^{(3,3)} (r_2) \right\rangle &= \frac{6}{Q^3} \left( \overbrace{\delta_{ad} \delta_{be} \delta_{cf} + \dots}^{6 \ {\rm terms}} - \frac{1}{Q-4} (\overbrace{ \delta_{ad} \delta_{be} + \dots}^{18 \ {\rm terms}}) \right. \notag \\
& \left. + \frac{2}{(Q-4)(Q-3)}(\underbrace{\delta_{ad}+ \dots}_{9 \ {\rm terms}}) - \frac{6}{(Q-4)(Q-3)(Q-2)} \right)  \PPPa. \label{t33t33}
\end{align}
In the last line we refrain from writing out all relevant permutations
of the tensor indices, but indicate only by a brace the total number
of terms, including the one shown explicitly.  We shall often use this
type of notation below.

\subsection{Two-point functions: general result}

Comparing the above results, a general pattern emerges that allows us
to write down the two-point function
\begin{equation*}
 \left\langle t^{(p,N)}_{a_1,a_2,\ldots,a_p}(r_1) t^{(p,N)}_{b_1,b_2,\ldots,b_p}(r_2) \right\rangle
\end{equation*}
for any number of spins $N$.

We first notice that the combination of Kronecker deltas factorizes
in the correlation functions and depends only on $p$. Defining
\be
 \delta^{(k,p)} = \overbrace{\delta_{a_1,b_1}\delta_{a_2,b_2}\cdots\delta_{a_k,b_k} + \dots}^{{p \choose k}^2 k! \ {\rm terms}}
\ee
the relevant combination is
\be
 \Delta^{(p)} = \sum_{k=0}^p \frac{(-1)^k k!}{(Q-2p+1+k)_k} \, \delta^{(k,p)} \,,
\ee
where we have defined the falling factorial $(x)_n = x(x-1)\cdots(x-(n-1))$.
Let us also define the following notation
\be
 {\mathbb P}^{(k,N)} = \PkN
\ee
for the probability of having $k$ propagating FK clusters linking the two groups
of $N$ spins; note that due to the internal permutations within each group of spins
this diagram corresponds to ${N \choose k}^2 k!$ terms.

The (conjectured) general two-point correlator can then be written
\be
 \left\langle t^{(p,N)}_{a_1,a_2,\ldots,a_p}(r_1) t^{(p,N)}_{b_1,b_2,\ldots,b_p}(r_2) \right\rangle = \Delta^{(p)} \sum_{k=p}^N \frac{(Q-2p)_{N-p} (Q-p-k)_{N-k} (k)_p}{Q^{2N-k}} \, {\mathbb P}^{(k,N)} \,.
\ee
The reader can verify that this expression reproduces (\ref{t01t01})--(\ref{t11t11}),
(\ref{t02t02})--(\ref{t22t22}) and (\ref{t03t03})--(\ref{t33t33}) as it should.

\subsection{Examples of three-point functions}
\label{secThreePointlattice}

The techniques deployed so far in this section can obviously be applied to the calculation of
correlation functions involving more than two operators. We shall here present some sample
results for three-point functions of operators that each act on two spins. In the corresponding
diagrams the three groups of each two spins are depicted on top, the lower left, and the lower right.
We imagine those groups of spins to be situated in small neighborhoods around
the points $r_1$, $r_2$ and $r_3$ respectively.

The result for the three-point function of the energy operator, defined in (\ref{energydef}), reads
\begin{multline}
\langle \epsilon (r_1) \epsilon (r_2) \epsilon (r_3)\rangle = \frac{(Q-2)(Q-1)}{Q^2} \PPb + \frac{Q-1}{Q} \PPa + \left(\frac{Q-1}{Q}\right)^2
\left[ \PPc \right. \\
\left. + \PPd - \Pdiff \left( \Pa(r_{12})+\Pa(r_{23})+\Pa(r_{31}) \right) \right] +  \left(\frac{Q-1}{Q}\right)^3
\left[ \PPh \right. \\
\left. + \PPe + \PPf +\PPg -  \Pdiff \left( \underbrace{ \Pb(r_{12})+\Pc(r_{12})+ \dots }_{3 \ {\rm coordinate \ permutations}}  \right) \right].
\label{eee}
\end{multline}
The meaning of the diagrams should by now be clear: each line denotes a Kronecker delta
identifying the values of the corresponding spins, and sums over permutations of the spins
inside each group are implied. Moreover, sums over permutations of the three groups of spins
are implied as well.

Let us illustrate those remarks on the first diagram appearing in (\ref{eee}). The first spin in the
$r_1$ group is linked to one of the two spins in either the $r_2$ or the $r_3$ group; this gives
four possibilities. The second spin in the $r_1$ group is then linked to one of the two spins in
remaining group, giving two possibilities. Finally the two remaining spins are linked in a unique
way. The first diagram thus corresponds to $8$ terms.

Similarly, in the second diagram in (\ref{eee}) the first spin in the $r_1$ group is linked to a
spin in each of the two other groups; this gives four possibilities. The three remaining spins are then
linked in a unique way. There are therefore $4$ terms corresponding to that diagram.
As a final example, the fourth diagram in (\ref{eee}) corresponds to $6$ terms (there are three
choices of the non-linked group, and two possible linkings of the two remaining groups).

Some of the terms in (\ref{eee}) factorize as products of probabilities for one of the groups, and
for the mutual linking of the remaining two groups. In that case a sum over the three choices of the
latter two groups is implied. When this happens, the coordinate choices for the latter two groups
are indicated (as $r_{12}$, $r_{23}$ or $r_{31}$).

We shall also need results for the mixed correlation functions of $\epsilon$ and $t^{(2,2)}$ operators.
The first of these vanishes (as we are used to by now):
\begin{equation}
\left\langle \epsilon (r_1) \epsilon (r_2)t_{ab}^{(2,2)} (r_3) \right\rangle = 0 \,.
\end{equation}
However, the other is (somewhat surprisingly) non-zero:
\begin{multline}
\left\langle \epsilon (r_1) t_{ab}^{(2,2)} (r_2) t_{cd}^{(2,2)} (r_3) \right\rangle = \frac{2}{Q^2} \left( \delta_{ac} \delta_{bd} + \delta_{ad} \delta_{bc} - \frac{1}{Q-2} (\delta_{ac} +\delta_{bd} + \delta_{ad}+ \delta_{bc})+ \frac{2}{(Q-2)(Q-1)} \right) \\
\times \left[ \PPa - \frac{1}{Q} \PPb + \frac{Q-1}{Q} \left(\PPcm+ \PPdm -\Pdiff \Pa(r_{23}) \right) \right].
\end{multline}
Obviously, in mixed correlation functions the groups of spins corresponding to different operators
do {\em not} play equivalent roles. Here the $\epsilon$ operator corresponds to the $r_1$ group,
shown in the top of the diagrams. When counting the number of terms implied by a given diagram,
only the $r_2$ and $r_3$ groups can be permuted; internal permutations of spins within any given
group are of course still allowed. In the diagrams where the three groups sustain different linkings,
we have emphasized this distinguishability by clearly labeling the $r_1$ group.

Finally, the correlation function of three $t^{(2,2)}$ operators is a sum of two diagrams:
\begin{equation}
\langle  t_{ab}^{(2,2)} (r_1) t_{cd}^{(2,2)} (r_2) t_{ef}^{(2,2)} (r_3)\rangle = \frac{G^{(1)}_{abcdef}(Q)}{Q^2} \PPa + \frac{G^{(2)}_{abcdef}(Q)}{Q^3} \PPb.
\end{equation}
with rather complicated prefactors that read
\begin{multline}
G^{(1)}_{abcdef}(Q) = \frac{16(2 Q-3)}{(Q-2)^3(Q-1)^2} - \frac{4(2 Q-3)}{(Q-2)^3(Q-1)} (\overbrace{ \delta_{ac} + \dots}^{12 \ {\rm terms}})  + 
   \frac{4}{(Q-2)(Q-1)}(\overbrace{\delta_{ac} \delta_{bd} + \dots}^{6 \ {\rm terms}})
   \\ + 
   \frac{2}{(Q-2)^2} (\underbrace{\delta_{ac} \delta_{de} + \dots}_{24 \ {\rm terms}}) +  
   \frac{4(Q-1)}{( Q-2)^3}(\underbrace{\delta_{ac} \delta_{ce} + \dots}_{8 \ {\rm terms}}) - 
   \frac{2}{Q-2} (\underbrace{\delta_{ac} \delta_{ce} \delta_{bd}  + \dots}_{24 \ {\rm terms}}) + 
   2 (\underbrace{\delta_{ac} \delta_{ce} \delta_{bd} \delta_{df} + \dots}_{4 \ {\rm terms}})
\end{multline}
and
\begin{multline}
G^{(1)}_{abcdef}(Q) = -\frac{8(Q^2 + Q - 4)}{(Q-2)^3(Q-1)^2} + \frac{2( Q^2 + Q - 4)}{(Q-2)^3(Q-1)} (\overbrace{ \delta_{ac} + \dots}^{12 \ {\rm terms}})  - 
   \frac{4}{(Q-2)(Q-1)}(\overbrace{\delta_{ac} \delta_{bd} + \dots}^{6 \ {\rm terms}})
   \\- 
   \frac{Q}{(Q-2)^2} (\underbrace{\delta_{ac} \delta_{de} + \dots}_{24 \ {\rm terms}})
    +(\underbrace{\delta_{ac} \delta_{de} \delta_{bf} + \dots}_{8 \ {\rm terms}})  -  
   \frac{2(3 Q- 4)}{( Q-2)^3}(\underbrace{\delta_{ac} \delta_{ce} + \dots}_{8 \ {\rm terms}}) + 
   \frac{2}{Q-2} (\underbrace{\delta_{ac} \delta_{ce} \delta_{bd}  + \dots}_{24 \ {\rm terms}}) \,.
\end{multline}

\subsection{Fusion}

Consider the product of two cluster operators, $t^{(p_1,N_1)}$ and $t^{(p_2,N_2)}$, of respective ranks $p_1$ and $p_2$, and acting on
$N_1$ and $N_2$ distinct spins respectively:
\be
 t^{(p_1,N_1)}_{a_1,a_2,\ldots,a_{p_1}}(\sigma_1,\sigma_2,\ldots,\sigma_{N_1}) \times
 t^{(p_2,N_2)}_{b_1,b_2,\ldots,b_{p_2}}(\sigma_{N_1+1},\sigma_{N_1+2},\ldots,\sigma_{N_1+N_2})
\ee
If we symmetrize this product over the $\frac{(N_1 + N_2)!}{N_1! \, N_2!}$ possible ways of assigning $N_1$ (resp.\ $N_2$) out of the total number of spins $N \equiv N_1 + N_2$ to the leftmost (resp.\ rightmost) operator, and similarly symmetrize over the $\frac{(p_1 + p_2)!}{p_1! \, p_2!}$ permutations of the tensor indices (dividing by a factor of $2$ if the two cluster operators are identical), then the result can be written as a sum over the operators $t^{(p,N)}$ acting on all spins. We call this decomposition the {\em fusion} of $t^{(p_1,N_1)}$ and $t^{(p_2,N_2)}$ and denote it as $t^{(p_1,N_1)} \otimes t^{(p_2,N_2)}$.
 
With $N_1 = N_2 = 1$ the only non-trivial fusion is the following
\begin{align}
 t_a^{(1,1)}(\sigma_1) t_b^{(1,1)}(\sigma_2) +  t_b^{(1,1)}(\sigma_1) t_a^{(1,1)}(\sigma_2) &=
 t_{ab}^{(2,2)}(\sigma_1,\sigma_2) + \frac{2}{Q(Q-2)}
 \left( t_a^{(1,2)}(\sigma_1,\sigma_2) + t_b^{(1,2)}(\sigma_1,\sigma_2) \right) \nonumber \\
 & + \frac{2}{Q^2(Q-1)} t^{(0,2)}(\sigma_1,\sigma_2) \,. \label{t11fusion}
\end{align}
It is convenient to rewrite it in a short-hand notation where we omit the spins and tensor indices, conserving only the number of symmetrized terms in the form of an integer inside a pair of parentheses surrounding the corresponding operator (on the right-hand side), or the fusion product (on the left-hand side). In this notation (\ref{t11fusion}) reads simply
\be
 \left( 2 t^{(1,1)} \otimes t^{(1,1)} \right) =
 \left( 1 t^{(2,2)} \right) + \frac{2}{Q(Q-2)} \left(2 t^{(1,2)} \right) + \frac{2}{Q^2(Q-1)} \left(1 t^{(0,2)} \right) \,.
\ee

Using (\ref{general_tensor}) extensively we have computed all non-trivial fusions for up to four spins.
For $N=3$ we find
\begin{align}
 \left( 9 t^{(2,2)} \otimes t^{(1,1)} \right) &=
 3 \left( 1 t^{(3,3)} \right) + \frac{8(Q-1)}{Q(Q-2)(Q-4)} \left( 3 t^{(2,3)} \right) +
 \frac{4 Q}{(Q-1)(Q-2)^2(Q-3)} \left( 3 t^{(1,3)} \right) \,, \\
 \left( 6 t^{(1,2)} \otimes t^{(1,1)} \right) &=
 2 \left( 1 t^{(2,3)} \right) + \frac{8}{Q(Q-2)} \left( 2 t^{(1,3)} \right) +
 \frac{12}{Q^2(Q-1)} \left( 1 t^{(0,3)} \right) \,.
\end{align}
For $N=4$ we have with $(N_1,N_2) = (3,1)$ 
\begin{align}
 \left( 16 t^{(3,3)} \otimes t^{(1,1)} \right) &=
 4 \left( 1 t^{(4,4)} \right) + \frac{12(Q-2)}{Q(Q-4)(Q-6)} \left( 4 t^{(3,4)} \right) +
 \frac{4 (Q-1)}{(Q-3)(Q-4)^2(Q-5)} \left( 6 t^{(2,4)} \right) \,, \\
 \left( 12 t^{(2,3)} \otimes t^{(1,1)} \right) &=
 3 \left( 1 t^{(3,4)} \right) + \frac{16(Q-1)}{Q(Q-2)(Q-4)} \left( 3 t^{(2,4)} \right) +
 \frac{12 Q}{(Q-1)(Q-2)^2(Q-3)} \left( 3 t^{(1,4)} \right) \,, \label{t23_fusion_t11} \\
 \left( 8 t^{(1,3)} \otimes t^{(1,1)} \right) &=
 2 \left( 1 t^{(2,4)} \right) + \frac{9}{Q(Q-2)} \left( 2 t^{(1,4)} \right) +
 \frac{24}{Q^2(Q-1)} \left( 1 t^{(0,4)} \right) \,, \label{t13_fusion_t11}
\end{align}
and finally with $(N_1,N_2) = (2,2)$ we find
\begin{align}
 \left( 18 t^{(2,2)} \otimes t^{(2,2)} \right) &=
 3 \left( 1 t^{(4,4)} \right) + \frac{12}{(Q-2)(Q-6)} \left( 4 t^{(3,4)} \right) +
 \frac{4(Q^2+9Q-16)}{(Q-1)(Q-2)^2(Q-4)(Q-5)} \left( 6 t^{(2,4)} \right) \,, \nonumber \\
 &+ \frac{72}{(Q-2)^3(Q-3)(Q-4)} \left( 4 t^{(1,4)} \right) + \frac{144}{Q(Q-1)^2(Q-2)^2(Q-3)} \left( 1 t^{(0,4)} \right) \,, \label{t22_fusion_t22} \\
 \left( 18 t^{(1,2)} \otimes t^{(2,2)} \right) &=
 3 \left( 1 t^{(3,4)} \right) + \frac{16(Q-1)}{Q(Q-2)(Q-4)} \left( 3 t^{(2,4)} \right) +
 \frac{12 Q}{(Q-1)(Q-2)^2(Q-3)} \left( 3 t^{(1,4)} \right) \,, \label{t12_fusion_t22} \\
 \left( 6 t^{(1,2)} \otimes t^{(1,2)} \right) &=
 2 \left( 1 t^{(2,4)} \right) + \frac{12}{Q(Q-2)} \left( 2 t^{(1,4)} \right) +
 \frac{24}{Q^2(Q-1)} \left( 1 t^{(0,4)} \right) \,. \label{t12_fusion_t12}
\end{align}

Several features in these expressions for $t^{(p_1,N_1)} \otimes t^{(p_2,N_2)}$ are worth pointing out.
First, the values of $p$ on the right-hand side invariably range from $|p_1-p_2|$ to $p_1+p_2$. This is
a familiar result in $d=2$, where it stems from the underlying ${\rm SU}(2)_q$ quantum group symmetry,
whereas here it is derived for general $d$ using only the generic $S_Q$ symmetry. It can also be
understood (albeit loosely) in the geometrical interpretation where the two operators on the left-hand
side insert respectively $p_1$ and $p_2$ FK clusters; after the fusion some of the clusters emanating
from one operator may (but need not) coalesce with, or annihilate, those coming from the other operator.

Second, the coefficients on the right-hand side have obviously the pole structure familiar from the
cluster operators themselves. More interestingly, that pole structure depends only on the values of
$p_1,p_2$ and not on $N_1,N_2$. For instance, the right-hand sides of (\ref{t23_fusion_t11}) and (\ref{t12_fusion_t22})
are identical, and those of (\ref{t13_fusion_t11}) and (\ref{t12_fusion_t12}) differ only in the numerical
coefficient of one of the terms.

Finally, and maybe most interestingly, the coefficients in front of the terms on the right-hand side sometimes
vanish for particular values of $Q$, such as $Q=0$, $Q=1$ and $Q=2$. (In the case of (\ref{t22_fusion_t22})
non-integer roots appear as well.) This implies that for particular $Q$-values there are truncations, or selection
rules, that prevent certain operators from appearing in the fusion product. Once again, this is well-known
from the quantum group analysis in $d=2$, but appears here as a consequence of the $S_Q$ symmetry in
general dimension.
 
\section{Continuum limit of the critical Potts model and physical interpretation of the operators $t^{(k,N)}_{a_1,\dots,a_k}$}
\label{sec:CFT}

Using solely the $S_Q$ symmetry of the Potts model, we have been able to classify a very large class of operators and compute their discrete correlation functions in terms of probabilities on an arbitrary graph. In the remainder of this paper, we shall apply these results to specific examples of physical relevance ($Q=1$ for percolation, $Q=2$ for Ising, $\dots$), and argue that, at the critical point, the analytic continuation of the $S_Q$ symmetry implies some `mixing' of the operators by the scale transformation generators. Before doing so, we comment on the continuum limit of the Potts model and discuss the physical meaning of the operators $t^{(k,N)}_{a_1,\dots,a_k}$ previously introduced.

\subsection{Landau field theory and critical point}

Let $\phi_a$ denote the coarse-grained order parameter of the Potts model obtained as the scaling limit of the lattice operator $t_{a}^{(1,1)}$. It has $Q-1$ components as it satisfies $\sum_{a=1}^{Q} \phi_a = 0$. In terms of this order parameter, one can write an effective Landau theory 
\begin{equation}
\label{eqLandauPotts}
S = \int {\rm d}^d r \left( \frac{1}{2} \sum_a (\partial_\mu \phi_a)^2 + \frac{m^2}{2} \sum_a \phi_a^2 + g \sum_a \phi_a^3 + \dots \right),
\end{equation}
with the crucial constraint $\sum_{a} \phi_a = 0$. For $Q < 2$, a perturbative RG analysis near the upper critical dimension $d_{\rm uc} = 6$ shows the existence of a non-trivial fixed point describing a second-order phase transition~\cite{Amit}. These are the fixed points that we shall describe in the following. For $Q=2$, $\sum_a \phi_a^n$ vanishes for any odd $n$ (and $n=3$ in particular) because of the $S_2=\mathbb{Z}_2$ symmetry of the Ising model. As a consequence, the upper critical dimension becomes $d_{\rm uc}=4$ and the critical point corresponds to the Wilson-Fisher fixed point. In $d=2$ dimensions, the Potts model has a second-order phase transition for $Q \leq 4$~\cite{BaxterPotts}. 

\subsection{Magnetization and cluster operators}

At the critical point, we expect the system to be described by a Conformal Field Theory (CFT). As argued in the previous sections, the scaling limit of the operators $t^{(N,N)}_{a_1,\dots,a_N}$ should correspond to the $N$-cluster operators for $N \geq 2$, while $t^{(1,1)}$ was identified to be the magnetization operator (order parameter), and $t^{(0,1)}$ to the identity. We will denote these cluster operators in the scaling limit\footnote{In the following, we will ignore the renormalization of the fields $\phi_{\rm CFT} = a^{- \Delta} \phi_{\rm lattice + \dots}$ where $a$ is the UV cutoff. For simplicity, we will use the same notations for lattice and CFT fields. We will also ignore for the moment subleading corrections, corresponding to less relevant operators with the same symmetry that shall be addressed in section~\ref{secSubleading}. } by $t^{(N,N)}$, and we will denote by $\mathbb{I}$ and $\phi_a$ the identity and magnetization operators, respectively. Let $\Delta_{N}$ be the corresponding cluster (or watermelon) critical exponents, and $\Delta_{\sigma}$ the dimension of the 
magnetization $\phi_a$. Using the symmetry results of the previous sections and conformal invariance, one can readily write the generic form of the correlation functions at the critical point. For instance, for the $2$-cluster  operator $t_{ab}^{(2,2)}$, with dimension $\Delta_2$, we find
\begin{equation}
\label{eqWatermelon2pointex}
\langle t_{ab}^{(2,2)} (r_1) t_{cd}^{(2,2)} (r_2)\rangle = A(Q) \left( \delta_{ac} \delta_{bd} + \delta_{ad} \delta_{bc} - \frac{1}{Q-2} (\delta_{ac} +\delta_{bd} + \delta_{ad}+ \delta_{bc})+ \frac{2}{(Q-2)(Q-1)} \right) r_{12}^{-2  \Delta_{2}},
\end{equation}
where the index structure is the same as in the discrete context, see eq.~\eqref{t22t22}, and $A(Q)$ is some unknown regular function of $Q$. The general form of CFT correlation functions for other cluster operators can be readily obtained in the same way.

\subsection{Subleading operators}
\label{secSubleading}

\begin{figure}[!t]
\centering
\includegraphics[width=0.75\textwidth]{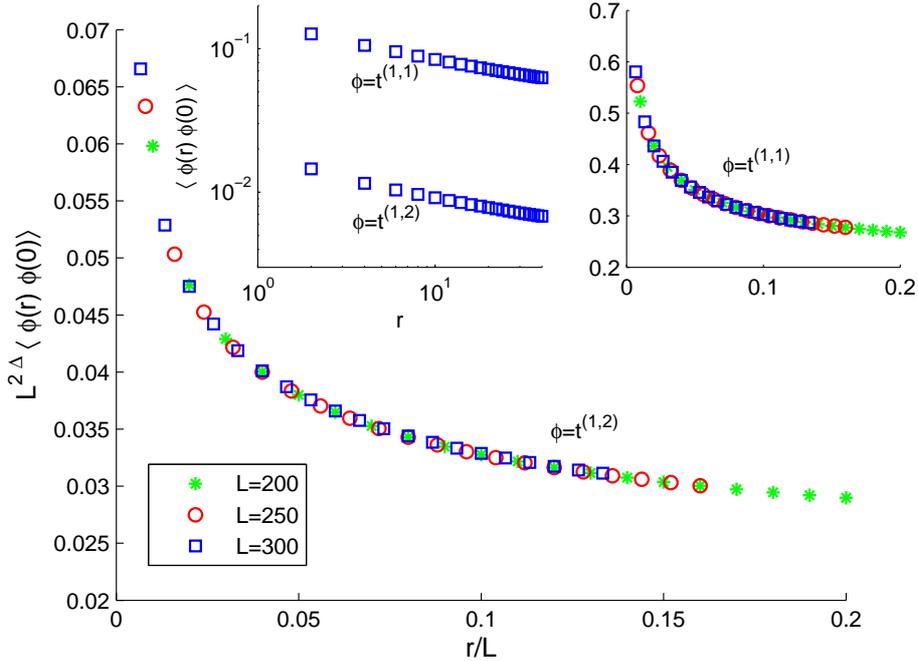}
\caption{Finite size scaling analysis of the two-point functions $\langle t^{(1,1)}(r) t^{(1,1)}(0) \rangle$ and  $\langle t^{(1,2)}(r) t^{(1,2)}(0)  \rangle$ in the two-dimensional 3-state Potts model on the square lattice. Data were collected on $\sim 10^3$ runs of $~\sim 10^7$ Monte Carlo steps each. As discussed in the text, because they transform under $S_Q$ in the same way, these lattice operators both scale with a leading contribution given by the magnetization exponent $\Delta=2h_{1/2,0}=\frac{2}{15}$. However, the lattice definition $t^{(1,2)}$ allows for a free parameter not fixed by symmetry (see eq.~\eqref{eqmoodified}) that could, in principle, be fine tuned to cancel this leading contribution and end up with a two-point function scaling with the critical exponent $\Delta^{(1)}_{\sigma}=2h_{3/2,0}=\frac{4}{3}$. }
\label{FigFSS}
\end{figure}

The physical meaning of the operators $t^{(k,N)}_{a_1,\dots,a_k}$, for $k<N$ is slightly more tricky. They actually correspond to {\it subleading} operators of the fields $t^{(k,k)}$, that is, less relevant operators with the same symmetry. We have already encountered one of these subleading operators, namely, the energy~\eqref{energydef}, which can be thought of as a subleading operator of the identity (vacuum) field. However, it is crucial to emphasize that in order to obtain the energy operator in~\eqref{energydef}, we had to subtract a constant term not fixed by symmetry in order to cancel the leading contribution coming from the identity channel. This is actually completely general, as the operator $t^{(k,N)}$ has exactly the same symmetry as $t^{(k,k)}$, so its two-point function will in general behave as a power-law with the same dominant critical exponent $\Delta_{k}$ (or $\Delta_\sigma$ for $k=1$), with subleading corrections that we will denote by $\Delta_{k}^{(p)}$, $p=1,2,\dots$ This relates to the well-known fact that lattice operators correspond to an infinite linear combination of scaling fields with the same symmetry. For instance, the operator
\begin{equation}
 t_{a}^{(1,2)}(\sigma_1,\sigma_2) = \delta_{\sigma_1 \neq \sigma_2} \left(\delta_{\sigma_1,a}+ \delta_{\sigma_2,a} - \frac{2}{Q}\right),
\end{equation}
transforms under $S_Q$ exactly as the order parameter $\phi_a$ (see Fig.~\ref{FigFSS}). Its two-point function at the critical point will therefore be dominated by a power-law with the same critical exponent $\Delta_\sigma$ as the magnetization, just like $t^{(0,N)}$ is dominated by the critical exponent of the identity $\Delta_0=0$ before the subtraction~\eqref{energydef}. However, it is in principle possible to define 
\begin{equation}
\label{eqmoodified}
 \tilde{t}_{a}^{(1,2)}(\sigma_1,\sigma_2) = \left( \delta_{\sigma_1 \neq \sigma_2} +K \right) \left(\delta_{\sigma_1,a}+ \delta_{\sigma_2,a} - \frac{2}{Q}\right),
\end{equation}
relaxing for the moment the constraint $\sigma_1 \neq \sigma_2$. The constant $K$ is not fixed by symmetry, and $ \tilde{t}_{a}^{(1,2)}$ has exactly the same properties as $t_{a}^{(1,2)}$. This constant $K$ could be fine-tuned to cancel the leading contribution $\Delta_\sigma$ to obtain the first subleading exponent $\Delta^{(1)}_\sigma$.  Similarly, the operators $t^{(1,N)}$ acting on $N \geq 3$ spins could be used to obtain other subleading exponents  $\Delta^{(N-1)}_\sigma$, where $\Delta_\sigma < \Delta^{(1)}_\sigma < \Delta^{(2)}_\sigma < \dots$ Unfortunately, this requires a lot of fine-tuning that cannot be fixed by representation theory, so in practical terms, the lattice operators $t^{(1,N)}$ will always be dominated by the leading magnetization critical exponent $\Delta_\sigma$.

In the following, we will denote by $\Delta^{(p)}_\sigma$ the subleading exponents corresponding to the magnetization operator, $\Delta^{(p)}_0$ those of the identity operator (with in particular, $\Delta^{(1)}_0$ the energy operator), and $\Delta^{(p)}_N$ those of the $N$-cluster operator.

\subsection{The two-dimensional case and exact critical exponents}

\label{secCFTexact}

In two dimensions, say on the square-lattice, the (outer and inner) hulls of the FK clusters constitutes a gas of loops~\cite{PottsLoop}. The in turn defines a height model (the loops being contour lines of the height) which can be argued to renormalize towards a Coulomb Gas (CG), that is, a compactified free boson with Lagrangian density $\mathcal{L}=\frac{g}{2\pi} (\nabla \phi)^2$, along with additional `electric charges' at infinity. The stiffness $g \in \left[2,4 \right]$ is given by $Q=2+2 \cos \frac{\pi g}{2}$, so that percolation has $g=\frac{8}{3}$. This leads to a Coulomb-Gas description of the Potts CFT in two dimensions~\cite{NienhuisLoop}. Using the more convenient parametrization $\sqrt{Q} = 2 \cos \frac{\pi}{x+1}$, the corresponding central charge is
\begin{equation}
c=1-\frac{6}{x(x+1)},
\end{equation}
and the corresponding critical exponents can be expressed nicely using the Kac formula
\begin{equation}
\label{eqKac}
h_{r,s} = \frac{\left(r(x+1)-s x \right)^2-1}{4 x (x+1)}.
\end{equation}
Using these notations, the magnetization (bulk) critical exponent reads $\Delta_\sigma= 2 h_{1/2,0}$, and the corresponding subleading exponents are $\Delta^{(p)}_\sigma= 2 h_{\frac{1}{2}+p,0}$~\cite{MagneticPotts,NienhuisLoop}. The identity operator has $\Delta_0=2 h_{1,1}$ while the corresponding subleading series is $\Delta^{(p)}_0=2 h_{p+1,1}$ -- in particular, the energy operator has dimension $\Delta^{(1)}_0=2 h_{2,1}$, recall also that subleading operators can be obtained by fusion with the energy  $\Phi_{2,1}$. The dimension of the $N$-cluster operators is also well-known to be $\Delta_N = 2 h_{0,N}$. Moreover, our analysis of the logarithmic structure (see below) led us to conjecture the following form for the subleading $N$-cluster exponents $\Delta^{(p)}_N = 2 (h_{p,N}+ p N)$.

\renewcommand{\arraystretch}{1.5}
\begin{table}
\begin{center}
\begin{turn}{90}
\begin{tabular}{|c|c|c|c|}
   \hline
   N & Operator & Physical meaning ($\dagger$) & Critical exponent in $d=2$ \\
   \hline
   1 & $t^{(0,1)}=1$ & identity & $\Delta_{0}=2 h_{1,1}=0$ \\
     & $t_a^{(1,1)}=\phi_a$ & magnetization & $\Delta_{\sigma}=2 h_{1/2,0}$ \\
   \hline
   2 & $t^{(0,2)}=\sum_{\substack{a,b \\ (a \neq b)}} :\phi_a \phi_b: \ = - \sum_{a} :\phi_a^2:$ & energy & $\Delta^{(1)}_{0}=2 h_{2,1}$ \\
    & $t_a^{(1,2)}=\sum_{b \neq a} :\phi_a \phi_b: - \frac{1}{Q} t^{(0,2)}$ & $1^{\rm st}$ subleading magnetization & $\Delta^{(1)}_{\sigma}=2 h_{3/2,0}$ \\  
    & $t_{ab}^{(2,2)}= :\phi_a \phi_b: - \frac{1}{Q-2}\left( t_a^{(1,2)}+ t_b^{(1,2)}\right)-\frac{1}{Q(Q-1)}t^{(0,2)}$ & 2-cluster & $\Delta_{2}=2 h_{0,2}$\\
   \hline
   3 & $t^{(0,3)}=\sum_{\substack{a,b,c \\ (\neq)}} :\phi_a \phi_b \phi_c:$ & $1^{\rm st}$ subleading energy & $\Delta^{(2)}_{0}=2 h_{3,1}$ \\
    & $t_a^{(1,3)}=\sum_{\substack{b \neq c \\ (\neq a)}} :\phi_a \phi_b \phi_c: - \frac{1}{Q} t^{(0,3)}$ & $2^{\rm nd}$ subleading magnetization & $\Delta^{(2)}_{\sigma}=2 h_{5/2,0}$ \\  
    & $t_{ab}^{(2,3)}= \sum_{c (\neq a,b)} :\phi_a \phi_b \phi_c: - \frac{1}{Q-2}\left( t_a^{(1,3)}+ t_b^{(1,3)}\right)-\frac{1}{Q(Q-1)}t^{(0,3)}$ & $1^{\rm st}$ subleading 2-cluster & $\Delta^{(1)}_{2}= 2 (h_{1,2}+2)$\\
    & $t_{abc}^{(3,3)}=  :\phi_a \phi_b \phi_c: - \frac{1}{Q-4}\left(t_{ab}^{(2,3)} + t_{ac}^{(2,3)}+t_{bc}^{(2,3)}\right)$ & 3-cluster & $\Delta_{3}= h_{0,3}$\\  
     & $-\frac{1}{(Q-2)(Q-3)} \left(t_a^{(1,3)}+t_b^{(1,3)}+t_c^{(1,3)} \right) -\frac{1}{Q(Q-1)(Q-2)}t^{(0,3)}$ &  & \\     
    \hline     
   4 & $t^{(0,4)}=\sum_{\substack{a,b,c,d \\ (\neq)}} :\phi_a \phi_b \phi_c \phi_d:$ & $2^{\rm st}$ subleading energy & $\Delta^{(3)}_{0}=2 h_{4,1}$ \\
    & $t_a^{(1,4)}=\sum_{\substack{b,c,d \\ (\neq a)}} :\phi_a \phi_b \phi_c \phi_d: - \frac{1}{Q} t^{(0,4)}$ & $3^{\rm rd}$ subleading magnetization & $\Delta^{(3)}_{\sigma}=2 h_{7/2,0}$ \\  
    & $t_{ab}^{(2,4)}= \sum_{c,d (\neq a,b)} :\phi_a \phi_b \phi_c \phi_d: - \frac{1}{Q-2}\left( t_a^{(1,4)}+ t_b^{(1,4)}\right)-\frac{1}{Q(Q-1)}t^{(0,4)}$ & $2^{\rm nd}$ subleading 2-cluster & $\Delta^{(2)}_{2}= 2 (h_{2,2}+4)$\\
    & $t_{abc}^{(3,4)}= \sum_{d (\neq a,b,c)}  :\phi_a \phi_b \phi_c \phi_d: - \frac{1}{Q-4}\left(t_{ab}^{(2,4)} + t_{ac}^{(2,4)}+t_{bc}^{(2,4)}\right)$ & $1^{\rm st}$ subleading 3-cluster & $\Delta^{(1)}_{3}= 2 (h_{1,3}+3)$\\  
     & $-\frac{1}{(Q-2)(Q-3)} \left(t_a^{(1,4)}+t_b^{(1,4)}+t_c^{(1,4)} \right) -\frac{1}{Q(Q-1)(Q-2)}t^{(0,4)}$ &  & \\   
     & $t_{abcd}^{(4,4)}= :\phi_a \phi_b \phi_c \phi_d: - \frac{1}{Q-6}\left(t_{abc}^{(3,4)} +  {\rm perm} \right) - \frac{1}{(Q-5)(Q-4)}\left(t_{ab}^{(2,4)} +  {\rm perm} \right)$ & 4-cluster & $\Delta_4=h_{0,4}$ \\
     & $  -\frac{1}{(Q-2)(Q-3)(Q-4)} \left(t_a^{(1,4)} + {\rm perm} \right) -\frac{1}{Q(Q-1)(Q-2)(Q-4)}t^{(0,4)}$ & &  \\     
    \hline         
  $\vdots$  &$\vdots$  & $\vdots$ & $\vdots$ \\
   \hline
\end{tabular}
\end{turn}
\end{center}

\caption{Classification of the scalar operators of the Potts Landau theory~\eqref{eqLandauPotts} in arbitrary dimension, and exact critical exponents in $d=2$ dimensions. Note that all the indices $a,b,c,d$ are pairwise different. ($\dagger$) Up to terms not fixed by symmetry (see text).}  
  \label{tab_opcontent}

\end{table}

\subsection{Operator content of the Potts Landau theory}

Although we have mostly focused on the lattice version of the Potts model so far, it is also possible to perform the very same analysis at the level of the Landau field theory~\eqref{eqLandauPotts}. In that case, there are no ambiguities between operators with the same symmetry, and there is a clear way to distinguish between, {\it e.g.}, the magnetization operator and its first subleading correction. On the other hand, it is easier to get a handle on the geometrical interpretation of the correlation functions on the lattice. 

The operator content of the field theory of the Potts model in $d$ dimensions is summarized in Tab.~\ref{tab_opcontent}. We also give the critical exponents of all these operators in the two-dimensional case where they can be computed using conformal invariance. Note that this classification only uses the $S_Q$ symmetry of the Potts model. Therefore, different operators transforming in the same way under $S_Q$ have in principle to be combined to form pure scaling fields, exactly as discussed on the lattice in Sec.~\ref{secSubleading}. For example, one might have to combine $t_a^{(1,2)}$ and $t_a^{(1,1)}$ in Tab.~\ref{tab_opcontent} -- this is allowed by symmetry -- to construct what one would like to call the subleading magnetization operator (see Sec.~\ref{secSubleading} for the lattice analog of this). This subtlety shall not be important to us and so we will mostly ignore it in the following.

\section{$Q \rightarrow 0$: spanning trees and spanning forests}
\label{sec:Qto0}

In this section, we discuss the limit $Q \rightarrow 0$ limit of the Potts model in relation
with spanning trees and forests. We show how our lattice correlation functions give
a direct geometrical interpretation of CFT correlators, and we comment on the appearance
of logarithms in the limit $Q \rightarrow 0$. In $d=2$, we reinterpret in our framework the well known
logarithmic partner of the identity operator, in relation with symplectic fermions
and resistor networks. Despite the fact that we will mostly deal with free (non-interacting)
theories in this section, we shall see in the following that the general ideas will work
in non-trivial, interacting cases as well.

\subsection{Spanning trees and dense polymers in $d=2$ dimensions}

\subsubsection{Spanning trees and lattice correlation functions}

Let us first discuss the case $d=2$. Recall the FK expansion of partition function
of the Potts model defined on a  graph G with $N$ sites~\eqref{ZFK}. 
Using the Euler relation $\left| A \right| +k(A) = N + \omega(A)$, where $\omega(A)$ is the number of loop in $A$, one can consider
the limit $Q \rightarrow 0$, $v \rightarrow 0$ with $w=\frac{Q}{v}$ fixed
\begin{equation}\label{eqPottsSpanningRescaled}
\lim \frac{1}{v^N} Z = \sum_{F} w^{k(F)} , 
\end{equation} 
where $F$ is a spanning forest of $G$, characterized by $\omega(F)=0$ (no loop).
In the limit $w \rightarrow 0$, only the so-called spanning trees $T$ of the graph $G$ survive up to a factor $w$, 
which are forests with only one connected component ($k(T)=1$)~\cite{Ivashkevich}. This model is critical,
as the critical line of the Potts model has a vertical tangent at the point $(Q,v)=(0,0)$ (see {\it e.g.}~\cite{JS2012}).
We will further consider the case of a square lattice in the following, for which 
the critical line is $v= \sqrt{Q}$.
To be more explicit, at the critical point, we have $v=w= \sqrt{Q}$, so that 
\begin{equation}
Z \underset{Q \rightarrow 0}{\sim} v^N w \mathcal{T}(G) = (\sqrt{Q})^{N+1} \mathcal{T}(G) , 
\end{equation} 
where $\mathcal{T}(G)$ is the number of spanning trees of the graph $G$.
This expression alternatively follows from the dense loop expansion of the Potts model at its critical point (see eq.~\eqref{eqZloop})
\begin{equation}
Z = Q^{N/2} \sum_{\rm loops} (\sqrt{Q})^{\rm number \ of \ loops} , 
\end{equation} 
where the loops are dual to the FK clusters. The limit $Q \rightarrow 0$ yields dense polymers (or dense 
self-avoiding walks), that are in one-to-one correspondence with spanning tree configurations.

In particular, we see that the partition function of the Potts model vanishes as $Q \rightarrow 0$.
In order to obtain non-trivial results, it is convenient to define the new correlators 
\begin{equation}
\langle \dots \rangle_0 = \frac{Z}{v^N} \langle \dots \rangle , 
\end{equation} 
so that the new critical partition function of the model is now $Z_0 \equiv \langle 1 \rangle_0 = \sqrt{Q} \mathcal{T}(G)$. 
Actually, because this limit $Q \rightarrow 0$ is a bit peculiar (the correlators are normalized in an unusual way), we will also have to rescale the observables in order to 
find non-trivial results\footnote{Of course, the global normalization factor of the operators defined in the previous sections
{\it is not} fixed by representation theory considerations.}. For example, we will define $\phi_{a} = \sqrt{Q} t^{(1,1)}_a = 
\sqrt{Q} \delta_{\sigma_i,a} - \frac{1}{\sqrt{Q}}$.  When $Q \rightarrow 0$, $\phi_{a}$ becomes
(formally) singular because of the factor $\frac{1}{\sqrt{Q}}$, and this will yield logarithms in the limit~\cite{CardyReplica,VJSPercoLogs}. Setting $\varphi_{a} = \sqrt{Q} \delta_{\sigma_i,a} $, it is not hard to check that all the correlation 
functions of this operator correspond to meaningful quantities. For example, since $ \langle \varphi_{a} \rangle = \frac{1}{\sqrt{Q}}$, we immediately find $ \langle \varphi_{a} \rangle_0 =  \mathcal{T}(G)$. It is also straightforward, using our results of Section~\ref{sec:discrete} to show that 
\begin{equation}
 \langle \varphi_{a}(r_i) \varphi_{b}(r_j) \rangle_0 \underset{Q \rightarrow 0}{\sim}  \frac{Z}{Q v^N} \left(1 - \cc \right) \sim {\sum_F}^{\prime} (\sqrt{Q})^{k(F)-2} \underset{Q \rightarrow 0}{\longrightarrow}\mathcal{T}(G \backslash \lbrace ij\rbrace) , 
\end{equation} 
where $\mathcal{T}(G \backslash \lbrace ij\rbrace)$ counts the number of spanning 2-tree forests, with one tree containing the site $i$ while the other contains $j$.
The symbol $\sum^{\prime}$ in the intermediate expression corresponds to a sum over forests configurations where $i$ and $j$ can belong to different trees.

\subsubsection{Continuum limit and logarithms}

In the field theory limit, we will denote $Z_0 = \sqrt{Q} A(Q)$, where $A(Q)$ is a regular function of $Q$ with a finite limit as $Q \rightarrow 0$. Obviously, in the scaling limit, we expect $\phi_{a}$ to become the magnetization operator, with critical exponent $\Delta_{\sigma}(Q)=2 h_{\frac{1}{2},0}$ (see Section~\ref{sec:CFT}), where we have used the standard Kac parametrization~\eqref{eqKac}. The exponent for the corresponding boundary operator would be $h_{1,3}$. Notice that $\Delta_{\sigma}(Q=0)=0$, so this operator will be mixed with the identity field at $Q=0$.
Using the discussion in section~\ref{sec:CFT}, we expect
\begin{equation}
 \langle \phi_{a}(r_i) \phi_{b}(r_j) \rangle_0 = \left( \delta_{a,b} \sqrt{Q} - \frac{1}{\sqrt{Q}}\right)  \tilde{A}(Q) r^{-2 \Delta_{\sigma}(Q)},
\end{equation} 
where $\tilde{A}(Q)$ is another regular function of $Q$ with a finite $Q \rightarrow 0$ limit.
The $Q \rightarrow 0$ limit of this equation is ill-defined, however, the correlation functions of $ \varphi_a = \phi_a+ 1/\sqrt{Q}$ have a finite
limit if one assumes that $A(0)=\tilde{A}(0)$%
\footnote{This assumption is strongly motivated by the fact that correlation functions would blow up at $Q=0$ if it did not hold. We also remark that at least in dimension $d=2$, all our results deduced from this are consistent with what is known from a more abstract study of LCFTs directly in the continuum limit (see~\cite{Symplecticfermions} in the context of $Q=0$).}
\begin{subequations}
\label{eqLogTrees}
\begin{eqnarray}
\left\langle \varphi_a(r)  \varphi_b(0)\right\rangle &=& 4 A(0) \left. \dfrac{\partial h_{\frac{1}{2},0}}{\partial \sqrt{Q}} \right|_{Q=0} \log r ,\\
\left\langle \varphi_a(r)\right\rangle &=& \left\langle \varphi_a(r) \mathbb{I}(0) \right\rangle  = A(0),
\end{eqnarray}
\end{subequations}
where we used $\mathbb{I}=1$ to denote the identity operator of the theory.
Using the expansion $h_{\frac{1}{2},0} = \frac{\sqrt{Q}}{4 \pi} + \dots$, we find that the ratio
\begin{equation}\label{eqSpanningTreesLog}
\dfrac{ \langle \varphi_a (r) \varphi_b (0) \rangle}{\left\langle \varphi_a(0)\right\rangle} = \dfrac{1}{\pi} \log r,
\end{equation} 
takes an universal form. This is probably the most simple correlation function showing a logarithm, and as we shall see,
it is in this case strongly related to the logarithmic form of the Green function of the Laplacian in 2D. 
Close to a boundary, we find
\begin{equation}\label{eqSpanningTreesLogBoundary}
\left. \dfrac{ \langle \varphi_a (r) \varphi_b (0) \rangle}{\left\langle \varphi_a(0)\right\rangle} \right|_{\rm boundary} = \dfrac{2}{\pi} \log r,
\end{equation} 
where we have used $h_{1,3} = \frac{\sqrt{Q}}{\pi} + \dots$

Let us come back to the bulk case to see that this logarithmic singularity in the correlation function $\langle \varphi_a (r) \varphi_b (0) \rangle$
should indeed be thought of as a logarithmic CFT feature. In order to do so, we analyze how all these fields transform under a scale transformation.
When $Q$ is generic, we have
\begin{subequations}
\begin{eqnarray}
 \mathbb{I}(\Lambda r) &=&  \mathbb{I}(r)  =  1,\\
  \phi_a(\Lambda r) &=& \Lambda^{-\Delta_{\sigma}(Q)}  \phi_a(r).
\end{eqnarray}
\end{subequations}
We can deduce from this the transformation law of $\varphi_a(r)$ when $Q=0$. We find
\begin{equation}
\varphi_a (\Lambda r) = 2 \left. \dfrac{\partial h_{\frac{1}{2},0}}{\partial \sqrt{Q}} \right|_{Q=0} \log \Lambda \mathbb{I}.
\end{equation} 
We see that the scale transformations generator is non-diagonalizable as it maps $\varphi_a$ onto the identity $\mathbb{I}=1$.
Using states instead of fields and Virasoro generators, this amounts to state that 
\begin{equation}
(L_0+\bar{L}_0) \Ket{\varphi_a} = -2\left. \dfrac{\partial h_{\frac{1}{2},0}}{\partial \sqrt{Q}} \right|_{Q=0} \Ket{\Omega}.
\end{equation} 
where $\Ket{\Omega}$ is the vacuum of the theory and $L_0$ and $\bar{L}_{0}$ are the usual Virasoro zero modes.

These universal logarithmic ratios have a very nice interpretation in terms of spanning trees (or equivalently, 
in terms of the dual dense polymers). 
Comparing our lattice expressions with eqs.~\eqref{eqLogTrees}, we find
\begin{equation}\label{eqSpanningTreesLog2}
\dfrac{\mathcal{T}(G \backslash \lbrace ij\rbrace)}{\mathcal{T}(G)} \sim \dfrac{1}{\pi} \log r_{ij},
\end{equation} 
which has a clear geometrical meaning. Note that the existence of logarithmic correlation functions for spanning trees is obviously not new (see {\it e.g.}~\cite{Ivashkevich,SpannLog1,SpannLog2}), but all the methods employed previously rely heavily on the free-fermionic/Laplacian property of the problem (see below), whereas our approach only used a simple-minded limit argument, that will turn out to apply also to highly non-trivial, interacting problems. It is also worth pointing out that these correlation functions {\it do not} appear directly in the abelian sand pile model~\cite{ASP}, as the latter only involves derivatives or differences of Green functions.

\subsubsection{Symplectic fermions}

To see how all this is related to a Laplacian problem, let us introduce Grassman variables ($\psi$, $\bar{\psi}$) on each site,
with the usual integration rules
\begin{equation}
\int {\rm d} \psi \ 1 = 0, \ \ \ \ \int {\rm d} \psi \ \psi = 1, \ \ \ \ \int {\rm d} \bar{\psi} \ 1 = 0, \ \ \ \ \int {\rm d} \bar{\psi} \ \bar{\psi} = 1.
\end{equation} 
Let us introduce the measure $\mathcal{D}\left[\psi,\bar{\psi}\right] = \prod_{i} {\rm d} \psi_i  {\rm d} \bar{\psi}_i $.
In terms of these fermionic variables, one can show that~\cite{ForrestsFermions}
\begin{subequations}
\begin{eqnarray}
\mathcal{T}(G) &=& \int \mathcal{D}\left[\psi,\bar{\psi}\right] \bar{\psi}_i \psi_i \mathrm{e}^{-S\left[\psi, \bar{\psi} \right]}, \\
\mathcal{T}(G\backslash \lbrace ij \rbrace) &=& \int \mathcal{D}\left[\psi,\bar{\psi}\right] \bar{\psi}_i \psi_i \bar{\psi}_j \psi_j  \mathrm{e}^{-S\left[\psi, \bar{\psi} \right]},
\end{eqnarray}
\end{subequations}
where the action reads
\begin{equation}
S\left[\psi, \bar{\psi} \right]  = - \sum_{i,j} \bar{\psi}_i L_{ij} \psi_j.
\end{equation} 
In this last equation, $L_{ij}$ are the matrix elements of the discrete Laplacian, with for the square lattice $L_{i,i}=4$
and $L_{i,j}=-1$ if $i$ and $j$ are neighbors. In the continuum limit, we expect this system to be described by a quantum
field theory with Lagrangian density
 \begin{equation}
 \label{eqSymplecticFermionL}
\mathcal{L}  = \frac{1}{4\pi} \partial_{\mu} \bar{\psi} \partial^{\mu} \psi.
\end{equation} 
This is of course the symplectic fermion theory~\cite{Symplecticfermions}, and the factor $1/4 \pi$ is conventional. 
The relation between the Potts model at $Q=0$ and this $c=-2$ (L)CFT is well-known
but it remains instructive to interpret how it results in this context.
The most striking feature of this theory is that its partition function vanishes exactly  
\begin{equation}
Z = \int \mathcal{D}\left[\psi(r),\bar{\psi}(r)\right] \mathrm{e}^{-\int d^2 x \mathcal{L}} = 0 ,
\end{equation} 
because of the zero mode of the Laplacian. The correlation functions are thus defined without the usual $Z$
factor
\begin{equation}
\langle \mathcal{O}\left[\psi, \bar{\psi} \right] \rangle \propto \int \mathcal{D}\left[\psi(r),\bar{\psi}(r)\right] \mathcal{O}\left[\psi, \bar{\psi} \right] \mathrm{e}^{-\int d^2 x \mathcal{L}}.
\end{equation} 
This parallels the normalization of lattice correlation functions.
This spectrum of this theory contains four rotation-invariant groundstates with scaling dimension $\Delta=0$ (in the CFT language, they have
conformal weights $(h,\bar{h})=(0,0)$). In addition to the two fields $\psi(r)$ and $\bar{\psi}(r)$,
there is an additional field $\omega(r) \equiv:\bar{\psi}(r)\psi(r):$ that is mixed with the vacuum $\Omega$ by the Hamiltonian $L_0+\bar{L}_0$.
It is possible to choose the normalization such that 
\begin{subequations}
\begin{eqnarray}
\langle \bar{\psi}(r) \psi(0)\rangle &=& 1, \\
\langle \omega(r) \omega(0)\rangle &=& \theta + 4 \log r,
\end{eqnarray}
\end{subequations}
where $\theta$ is a (non-universal) constant. The logarithmic divergence of the correlation function~\eqref{eqSpanningTreesLog}
therefore corresponds to the Jordan cell mixing the vacuum and the field $\omega(r) = :\bar{\psi}(r)\psi(r):$,
that we can thus identify, up to a multiplicative constant, with $\varphi_a(r)$.
The factor $4$ in front of the logarithm is of course related to our choice of normalization.

\subsubsection{Applications to resistor networks}

\begin{figure}[!t]
\centering
\includegraphics[width=0.8\textwidth]{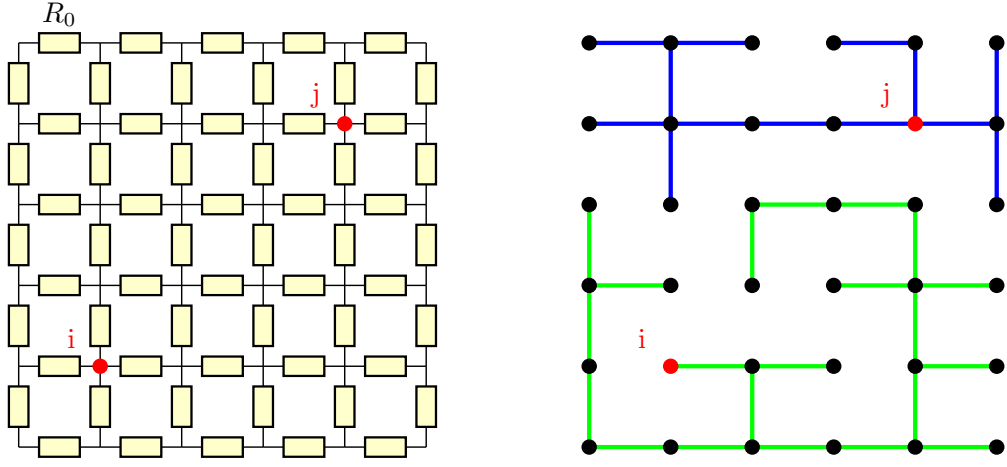}
\caption{The equivalent resistance $R_{\rm eq}(n,m)$ between the points $i$ and $j$ is given by $R_0 \mathcal{T}(G)/\mathcal{T}(G \backslash \lbrace ij\rbrace)$, where $\mathcal{T}(G \backslash \lbrace ij\rbrace)$ counts the number of spanning 2-tree forests, with one tree containing the site $i$ while the other covers $j$.}
\label{FigResistors}
\end{figure}

To conclude this section about spanning trees, let us discuss our result in the context of resistor networks. 
The problem is very simple: Let us define an infinite square network of resistors, where each bond carries
a resistance $R_0$. One would like to compute the equivalent resistance between two arbitrary points $i$ and $j$ of the network. 
The answer to this question is very simple and uses the Green's function of the Laplacian $\Delta (\log r) = 2 \pi \delta(r)$.
It is actually straightforward to perform the calculation directly on the square lattice. To measure the equivalent resistance  between $n$ and $m$,
let us consider a current $I$ flowing from $n$ to $m$. Using Kirchhoff's laws, one finds that the electric potential
satisfies the following Laplace equation
\begin{equation}
\Delta_{\rm discrete} V_r  \equiv \sum_{r' \ {\rm n.n.} \ r} V_{r'} - 4 V_r = R_0 I (\delta_{r,m}-\delta_{r,n}).
\end{equation} 
The equivalent resistance  $R_{\rm eq}(n,m)$ between $n$ and $m$ is then simply given by $R_{\rm eq}(n,m)=(V_n - V_m)/I$.
Using the lattice Green function of the Laplacian, it is straightforward to show that (see {\it e.g.}~\cite{Wu,Cserti}) 
\begin{equation}
R_{\rm eq}(n,m) = R_0 \int_{-\pi}^{\pi} \frac{\mathrm{d}x}{2 \pi} \int_{-\pi}^{\pi} \frac{\mathrm{d}y}{2 \pi} \dfrac{1-\cos(nx+my)}{2- \cos x - \cos y}.
\end{equation} 
The asymptotic behavior of this function yields the resistance between two arbitrary points separated by a distance $r$ 
\begin{equation}
\frac{R_{\rm eq}(r)}{R_0} = \frac{2 \gamma + 3 \log 2 }{2 \pi} + \frac{1}{\pi} \log \frac{r}{a}  + \mathcal{O}(r^{-2}),
\end{equation} 
where $a$ is a UV cutoff. The first term of this equation is a non-universal constant that depends on the square lattice structure,
however, the following behavior is universal
\begin{equation}
R_{\rm eq}(r) \sim \frac{R_0}{\pi} \log r,
\end{equation} 
and can be found directly from the continuum Green function $G_0(r)=\frac{1}{2 \pi} \log r$.
This result is directly related to eq.~\eqref{eqSpanningTreesLog} and eq.~\eqref{eqSpanningTreesLog2} using Kirchhoff theorem~\cite{Kirchhoff}:
the conductance $G(i,j)$ between two arbitrary points $i$ and $j$ of a network $G$ of resistors $R_0=1$ is given in terms
of spanning trees by the ratio $\mathcal{T}(G)/\mathcal{T}(G \backslash \lbrace ij\rbrace)$ (see Fig.~\ref{FigResistors}). The logarithmic divergence of eq.~\eqref{eqSpanningTreesLog}
is thus directly related to the simple conductance calculation of an infinite resistor network. What is more interesting, though, 
is the interpretation of the prefactor $1/\pi$ as a derivative of a critical exponent of the Potts model. Moreover, this approach using
a limit of the Potts model is quite general and not restricted to Laplacian problems, and we shall see several highly non-trivial
examples in the following. To conclude, let us also mention that the boundary result~\eqref{eqSpanningTreesLogBoundary}
has also a nice interpretation in terms of resistor networks. Indeed, let us consider a resistor network covering the upper half-plane $y\geq 0$,
and let us compute the resistance $R_{\rm eq}(r)$ between two points lying at the boundary $y=0$.
The Green function $G_0(x_0,y_0,x,y)$ of the Laplacian must satisfy in this case Neumann boundary conditions $\partial_y G_0(x_0,y_0,x,y=0)=0$.
We find $G_0(x_0,y_0=0,x,y=0)=\frac{1}{\pi} \log (x-x_0)$, so there is an additional factor $2$ as compared to the bulk case.
Therefore, the boundary resistance reads
\begin{equation}
\left. R_{\rm eq}(r) \right|_{\rm boundary} \sim \frac{2 R_0}{\pi} \log r,
\end{equation} 
in agreement with eq.~\eqref{eqSpanningTreesLogBoundary}.

\subsection{Spanning forests in $d=3$}
\label{secQ0d3}

We conclude this section on $Q=0$ by discussing how our results can be generalized to higher dimensions $d$ smaller
than the upper critical dimension $d_{\rm uc}=6$.
Equation~\eqref{eqPottsSpanningRescaled} remains valid in that case, the only difference being that
the model is believed to be critical for a finite non zero value $w_c$ of $w$~\cite{Forests3D}, meaning
that $v \propto Q$ as $Q \rightarrow 0$, instead of $v \propto \sqrt{Q}$ for $d=2$.
One ends up with spanning forests with a non-zero fugacity $w$, which
can be described in terms of an {\it  interacting} fermionic field theory~\cite{ForrestsFermions}
\begin{equation}
S  =  \int {\rm d}^d x \left( \partial_{\mu} \bar{\psi} \partial^{\mu} \psi + \frac{g}{2} \left[\partial_{\mu}(\bar{\psi}\psi) \right]^2 - g \bar{\psi} \psi \right),
\end{equation} 
with bare coupling $g \propto w$. To all orders of perturbation theory, this interacting fermionic field theory can be mapped onto a $\sigma$ model with OSp$(1|2)$ supersymmetry~\cite{ForrestsFermions}, or equivalently, to an O$(n)$-invariant $\sigma$ model analytically continued to $n=-1$. In two dimensions, these models are
(perturbatively) asymptotically free, with $\beta$ function
 \begin{equation}
\frac{{\rm d} g}{{\rm d} \log L} = \frac{3}{2 \pi} g^2 + \dots
\end{equation} 
The $g=0$ ($w=0$) fixed point then corresponds to the free symplectic fermion (spanning trees) theory~\eqref{eqSymplecticFermionL} -- up to normalization. 

It is easy to see that because $v \propto Q$, the Jordan cell for the identity operator that we found for $d=2$
does not appear in higher dimensions. This is because we have to normalize our operators with a factor $v$ in order
to find non-trivial results, so that the fact that $v \propto Q$ instead of $v \propto \sqrt{Q}$ will actually give way to
a cancellation of the divergences that we encountered for $d=2$. One can also notice that the scaling dimension
of the magnetization vanishes only for $d=2$. At one-loop for example, it reads~\cite{Amit}
 \begin{equation}
\Delta_{\phi} = \frac{\eta +d-2}{2}=2- \frac{5 Q -16}{9Q-30}\epsilon + \mathcal{O}(\epsilon^2), 
\end{equation} 
with $\epsilon = 6-d$. For $Q=0$, $\Delta_{\phi}$ decreases as a function of $\epsilon$ and reaches $0$ {\it only}
for $d=2$ (This is of course only imperfectly brought out by the one-loop result that reads $\Delta_\phi = -2/15 + \dots$ for $d=2$). However, we have considered only the simplest Jordan cell at $Q=0$, and we do not exclude the possibility
of a logarithmic structure for more complicated observables in $d=3$. We leave the study of such observables
at $Q=0$ for future work.

\section{$Q \rightarrow 1$: Bond percolation in $d$ dimensions}

In this section, we revisit using our framework the results of~\cite{VJSPercoLogs} regarding the mixing of the energy operator with the $2$-cluster operator in percolation ($Q=1$). Both two- and three-point functions are addressed. 

\subsection{$Q \rightarrow 1$ limit and logarithmic observable in percolation}

\begin{figure}[!t]
\centering
\includegraphics[width=0.4\textwidth]{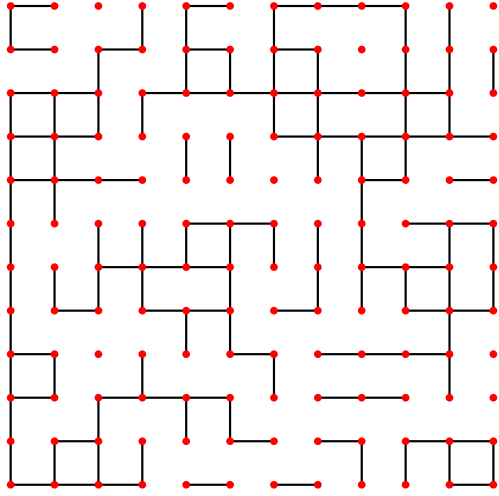}
\caption{Bond percolation configuration in two dimensions.}
\label{FigPerco}
\end{figure}

The limit $Q \to 1$ of the Potts model describes the bond-percolation problem (see Fig.~\ref{FigPerco} for an example of configuration in two dimensions). Let us first study the operators acting on two spins from a quantum field theory point of view. Recall that the energy operator is given by $\varepsilon (r_i) \equiv t^{(0,2)}(r_i) - \langle t^{(0,2)} \rangle$, where we subtracted the bulk expectation value of $\varepsilon$ so as to obtain a well-defined scaling field. Its two-point function is given by
\begin{equation}
\displaystyle \langle \varepsilon (r)  \varepsilon (0) \rangle = \tilde{A}(Q) (Q-1) r^{-2 \Delta_{\varepsilon}(Q)} \,,
\end{equation} 
where $\tilde{A}(Q)$ is a regular function of $Q$, with a finite non-zero limit $\tilde{A}(1)$ for $Q \to 1$, and $\Delta_{\varepsilon}(Q) = \Delta_{0}^{(1)}$. The reasons why $\langle \varepsilon (r)  \varepsilon (0) \rangle$ should vanish at $Q=1$ is obvious from a lattice point of view, since percolation is a problem of {\em uncorrelated} bonds. On the other hand, we have already argued that $t^{(2,2)} $ corresponds to the 2-cluster operator, so that its two-point function reads  
\begin{multline}
\label{eqTwoPointpsiHat}
\langle t^{(2,2)}_{ab} (r) t^{(2,2)}_{cd} (0) \rangle = \frac{2 A(Q)}{Q^2} \left( \delta_{ac} \delta_{bd} + \delta_{ad} \delta_{bc} - \frac{1}{Q-2} \left(\delta_{ac}+\delta_{ad}+\delta_{bc}+\delta_{bd} \right)  \right. \\
+\left. \frac{2}{(Q-1)(Q-2)} \right)
 \times r^{-2 \Delta_{2}},
\end{multline} 
where $A(Q)$ is again a regular function of $Q$ when $Q\rightarrow 1$, and the factor $2/Q^2$ is purely conventional. The rest of the correlation function is fixed by representation theory\footnote{Note that we have slightly changed the definition of the function $A(Q)$ with respect to~\eqref{eqWatermelon2pointex}.}.

In the formal limit $Q\rightarrow 1$ the two-point function~(\ref{eqTwoPointpsiHat}) diverges. To cure this, we introduce a new field
\begin{equation}
 \psi_{ab}  (r) = t^{(2,2)}_{ab}  (r) + \frac{2}{Q(Q-1)}  \varepsilon (r) \,,
 \label{eq_tilde_psi}
\end{equation}
mixing the energy and $2$-cluster operators of the Potts model. Its two-point function is easily computed and in order to have a finite $Q \to 1$ limit, we must require $A(1)=\tilde{A}(1)$, and that $\Delta_{\varepsilon}=\Delta_{2}$ at $Q=1$~\cite{VJSPercoLogs}. This implies that the fractal dimension $d_{\rm RB}$ of the so-called ``red bonds'' (also called ``cutting bonds'') is related to the thermal exponent $\nu$ via $d_{\rm RB} = \nu^{-1}$. This is indeed a well-known percolation result~\cite{Coniglio}, valid in any dimension. We find
\begin{equation}
\label{eq_PercoLog}
 \langle \psi_{ab}(r) \psi_{cd}  (0) \rangle = 2 A(1) r^{- 2 \Delta_{4}} \left[
 \left( \delta_{ac}+\delta_{ad}+\delta_{bc}+\delta_{bd} + \delta_{ac} \delta_{bd} + \delta_{ad} \delta_{bc} \right) +
 2 \delta \log r \right] \,,
\end{equation} 
where we have defined
\begin{equation}
\label{eqLimitbeta}
 \delta \equiv \ 2  \times \lim_{Q \rightarrow 1} \frac{\Delta_{2} - \Delta_{\varepsilon}}{Q-1}  \, ,
\end{equation} 
and $\delta=\frac{2 \sqrt{3}}{\pi}$ in $d=2$ dimensions. Under a scale transformation $r \to \Lambda r$, it is straightforward to show that $\psi_{ab}(r)$ transforms as
\begin{equation}
\psi_{ab}(\Lambda r) = \Lambda^{-\Delta_2} \left(\psi_{ab}(r)  + \delta \log \Lambda \ \varepsilon (r) \right), 
\end{equation} 
so that the scale transformation generator is not diagonalizable as it mixes $\psi_{ab}$ and $\varepsilon$ into a Jordan cell.

To understand what this logarithmic correlation means on the lattice, we define $\psi_{ab}  (r_i) \equiv t^{(2,2)} (r_i) + \frac{2}{Q(Q-1)}  \varepsilon (r_i)$ which is the lattice version of~\eqref{eq_tilde_psi}, where we recall that $\varepsilon (r_i)$ is defined as in the continuum limit, that is, subtracting from $t^{(0,2)}$ its expectation value $\frac{Q-1}{Q} \Pdiff$.

Using the results of section~\ref{sec:discrete}, it is straightforward to compute the lattice correlation function of $\psi_{ab}$ in terms of FK probabilities. We find a well-defined $Q  \to 1$ limit
\begin{multline}
\label{eqpercolattice}
 \displaystyle \langle \psi_{ab} (r_1) \psi_{cd} (r_2) \rangle =  
 2 \left( \delta_{ac}+\delta_{ad}+\delta_{bc}+\delta_{bd} + \delta_{ac} \delta_{bd} +
 \delta_{ad} \delta_{bc} \right) \Pa \\
 + 4 \left[\Pc+\Pb - 2 \Pa  - \Pdiff^2 \right]  .
\end{multline}  
Comparing the CFT~\eqref{eq_PercoLog} and the lattice~\eqref{eqpercolattice} results, we first obtain that 
\begin{equation}
\Pa \sim A(1)  r^{-2 \Delta_2},
\end{equation}
as was of course expected from the definition of the 2-cluster operator. More interestingly, the logarithmic term in~\eqref{eq_PercoLog} can be identified with~\cite{VJSPercoLogs}
\begin{equation}
\Pc+\Pb - \Pdiff^2 \sim A(1) \delta r^{-2 \Delta_2} \log r.
\end{equation}
This prediction was verified numerically in two-dimension in~\cite{VJSPercoLogs}. Let us emphasize however that the same prediction is expected to hold in higher dimensions, the only difference being the value of the critical exponent $\Delta_2$ and of the universal number $\delta$.

\subsection{Logarithmic $3$-point functions}

The fact that the operator $\psi_{ab}  (r_i) \equiv t^{(2,2)} (r_i) + \frac{2}{Q(Q-1)}  \varepsilon (r_i)$ is well-defined in the percolation limit $Q \to 1$ can also be checked at the level of $3$-point correlation functions. At the critical point, the 3-point functions of the fields $\epsilon$ and $t_{ab}^{(2,2)}$ are completely fixed (up to a constant) by conformal invariance and the $S_Q$ representation theory. Using the lattice results of section~\ref{secThreePointlattice}, we obtain, for example for the energy operator 
\begin{equation}
\langle \epsilon (r_1) \epsilon (r_2) \epsilon (r_3)\rangle = \frac{(Q-1)\alpha(Q)}{(r_{12} r_{23} r_{13})^{\Delta_{\varepsilon}}}.
\end{equation}
We will also need the following correlation functions
\begin{multline}
\langle \epsilon (r_1) t_{ab}^{(2,2)} (r_2) t_{cd}^{(2,2)} (r_3)\rangle = \beta(Q) \left( \delta_{ac} \delta_{bd} + \delta_{ad} \delta_{bc} - \frac{1}{Q-2} (\delta_{ac} +\delta_{bd} + \delta_{ad}+ \delta_{bc})+ \frac{2}{(Q-2)(Q-1)} \right) \\
\times \frac{1}{(r_{12} r_{13})^{\Delta_{\varepsilon}} r_{23}^{2 \Delta_{2} - \Delta_{\varepsilon}}}.
\end{multline}
\begin{equation}
\langle  t_{ab}^{(2,2)} (r_1) t_{cd}^{(2,2)} (r_2) t_{ef}^{(2,2)} (r_3)\rangle = \left(\frac{G^{(1)}_{abcdef}(Q)}{Q^2} F_{1}(Q) + \frac{G^{(2)}_{abcdef}(Q)}{Q^3} F_{2}(Q) \right) \times \frac{1}{(r_{12} r_{23} r_{13})^{\Delta_{2}}}  .
\end{equation}
Note also that our lattice results indicate that some of the 3-point functions vanish, for instance
\begin{equation}
\langle \epsilon (r_1) \epsilon (r_2)t_{ab}^{(2,2)} (r_3)\rangle = 0 ,
\end{equation}
thus indicating that the $S_Q$ symmetry strongly constrains the fusion rules of the Potts CFT even in dimension $d>2$.

Introducing the new field $\psi_{ab} $ and taking the limit $Q \to 1$, we find that the correlation functions are well-defined provided that
\begin{align}
2 \alpha(1) &= \beta(1), \notag \\
 \beta(1) &= 2 (F_1(1) - F_2(1)), \notag \\
 \alpha^\prime(1) &=F_1(1)+\frac{3}{2} \beta^\prime(1)+2(F^\prime_1(1) - F^\prime_2(1)).
\end{align}
At the percolation point ($Q=1$), we have
\begin{equation}
\langle \epsilon (r_1) \epsilon (r_2) \psi_{ab} (r_3)\rangle = \frac{2 (F_1(1) - F_2(1)}{(r_{12} r_{23} r_{13})^{\Delta}},
\end{equation}
\begin{equation}
\langle \epsilon (r_1) \psi_{ab} (r_2) \psi_{cd} (r_3)\rangle = \frac{2 (F_1(1) - F_2(1))}{(r_{12} r_{23} r_{13})^{\Delta}} \left(\delta_{ac}\delta_{bd}+\delta_{ad}\delta_{bc} +\delta_{ac}+\delta_{bd}+\delta_{ad}+\delta_{bc} + 4 \delta \log \left( \frac{r_{23}}{a}\right) \right),
\end{equation}
and
\begin{multline}
\langle \psi_{ab} (r_1) \psi_{cd} (r_2) \psi_{ef} (r_3)\rangle = \frac{1}{(r_{12} r_{23} r_{13})^{\Delta}} \left[ 2 F_{1}(1)(\delta_{ac}\delta_{ce}\delta_{bd}\delta_{df} + \dots)+2 (F_{1}(1)-F_{2}(1)) (\delta_{ac}\delta_{ce}\delta_{bd} + \dots) \right.\\
\left. - 2 F_2(1) (\delta_{ac}\delta_{ce} + \dots) + F_2(1) (\delta_{ac}\delta_{de}\delta_{fb} + \dots) + (2 F_1(1) - F_2(1))(\delta_{ac}\delta_{be} + \dots) 
\right.\\
\left.+ 4 (F_1(1) - F_2(1)) \left(\delta_{ac}\delta_{bd}+\delta_{ad}\delta_{bc} +\delta_{ac}+\delta_{bd}+\delta_{ad}+\delta_{bc}\right) \delta \log \left( \frac{r_{23}r_{13}}{r_{12} a}  \right) \right.\\
\left.+ 4 (F_1(1) - F_2(1)) \left(\delta_{ae}\delta_{bf}+\delta_{af}\delta_{be} +\delta_{ae}+\delta_{bf}+\delta_{af}+\delta_{be}\right) \delta \log \left( \frac{r_{12}r_{23}}{r_{13} a}  \right) \right.\\
\left.+ 4 (F_1(1) - F_2(1)) \left(\delta_{ce}\delta_{df}+\delta_{cf}\delta_{de} +\delta_{ce}+\delta_{df}+\delta_{cf}+\delta_{de}\right) \delta \log \left( \frac{r_{12}r_{13}}{r_{23} a}  \right) \right.\\
\left.
+ 8 (F_1(1) - F_2(1)) \left( {\rm cst} - \delta^2 \log \left( \frac{r_{12} r_{23} r_{31}}{a^3} \right)^2 \right)\right].
\end{multline}
We have introduced a UV cutoff $a$ in order to make the argument of the logarithms dimensionless. This also makes more transparent which terms cannot be universal in these expressions. Recall also that the universal number $\delta$ is given by~\eqref{eqLimitbeta}.

The very same correlation functions can be expressed on the lattice in terms of FK probabilities. Some results relevant to our purpose are gathered in appendix~\ref{app3pointPerco}. By carefully comparing the CFT and lattice correlation functions, we first see that
\begin{equation}
\PPa \sim \frac{F_1(1)}{(r_{12} r_{23} r_{31})^{\Delta_{2}}},
\qquad \qquad
\PPb \sim \frac{F_2(1)}{(r_{12} r_{23} r_{31})^{\Delta_{2}}},
\end{equation}
which was expected from the definition of the $2$-cluster operator. We also find the following logarithmic probabilities
\begin{equation}
\PPcmm + \PPdmm - \Pdiff \Pa(r_{13}) \sim  \frac{F_1(1)-F_2(1)}{(r_{12} r_{23} r_{31})^{\Delta_2}} \left({\rm cst} + \delta \log \left( \frac{r_{12} r_{23}}{r_{13}}\right) \right)
\end{equation}
\begin{multline}
\PPh +
\PPe +
\PPf +
\PPg -
\Pdiff \left[ \Pb(r_{12}) + \Pc(r_{12}) + \dots \right] + 2 \Pdiff^3  \\
\sim \frac{F_1(1)-F_2(1)}{(r_{12} r_{23} r_{31})^{\Delta_2}}
\left[ {\rm cst} - \delta^2 \log \left(r_{12} r_{23} r_{31} \right)^2 + {\rm cst} \times \log \left(r_{12} r_{23} r_{31} \right)  \right]. 
\end{multline}
As a consequence of the previous equations, we find
\begin{equation}
\PPcmm + \PPdmm - \Pdiff \Pa(r_{13}) +  3 \leftrightarrow 2 \sim  \frac{F_1(1)-F_2(1)}{(r_{12} r_{23} r_{31})^{\Delta_2}} \left({\rm cst} + 2 \delta \log  r_{23} \right),
\end{equation}
which is also consistent with the correlation function results. Let us emphasize again that these results are expected to be true in any dimension below the upper critical dimension $d_{\rm uc}=6$ of the percolation problem. It would be very interesting to verify these predictions numerically.

\section{$Q \rightarrow 2$: Ising model}

As can be seen from~\eqref{eqTwoPointpsiHat}, the $Q \to 2$ limit (FK formulation of the Ising model) is also ill-defined. This is because the 2-cluster operator $t^{(2,2)}_{ab}  (r)$ is mixed with the `vector' operator $t^{(1,2)}_{a}  (r)$ at $Q=2$. From the point of view of critical exponents in two dimensions, this is consistent with the fact that $h_{0,2}=h_{3/2,0}$ at $Q=2$, where $h_{3/2,0}$ is the dimension of the first subleading magnetization operator. Unfortunately, as discussed extensively in section~\ref{sec:CFT}, although the mixing is clear in the continuum (it involves the 2-cluster and the subleading magnetization operators), the situation is more intricate on the lattice as the magnetization and the subleading magnetization operators have the same symmetry, so it is hard to construct a precise lattice version of the subleading spin field. In practical terms, the two-point function of $t^{(1,2)}_{a}  (r)$ defined in~\eqref{tab22} has a dominant contribution given by the magnetization operator. 

This means that logarithmic corrections will affect only the first subleading power-law. Let $\delta\equiv 2 \lim_{Q \to 2} \frac{\Delta_{\sigma}^{(1)}-\Delta_2}{Q-2}$, so that we have $\delta=\frac{2}{\pi}$ in $d=2$ dimensions. In terms of probabilities, we find that the mixing at $Q \to 2$ implies that
\begin{align}
 \Pa & \sim A \times r^{-\Delta_{2}} + \dots ,\\
 \Pb & \sim B \times r^{-\Delta_{\sigma}} + 4 \delta \times A \times r^{-\Delta^{(1)}_{\sigma}} \log r + \dots
\end{align}
with $\Delta_{2}=\Delta_{\sigma}^{(1)}$. We thus see that the logarithmic corrections arises only in the subleading term, for the reason explained above. Unfortunately, there is {\it a priori} no simple way using symmetry to form a linear combination of probabilities to get rid of the dominant term $B r^{-\Delta_{\sigma}}$, making this prediction very hard to test numerically.

\section{General pattern of the logarithmic structure}

\begin{figure}[!t]
\centering
\includegraphics[width=0.8\textwidth]{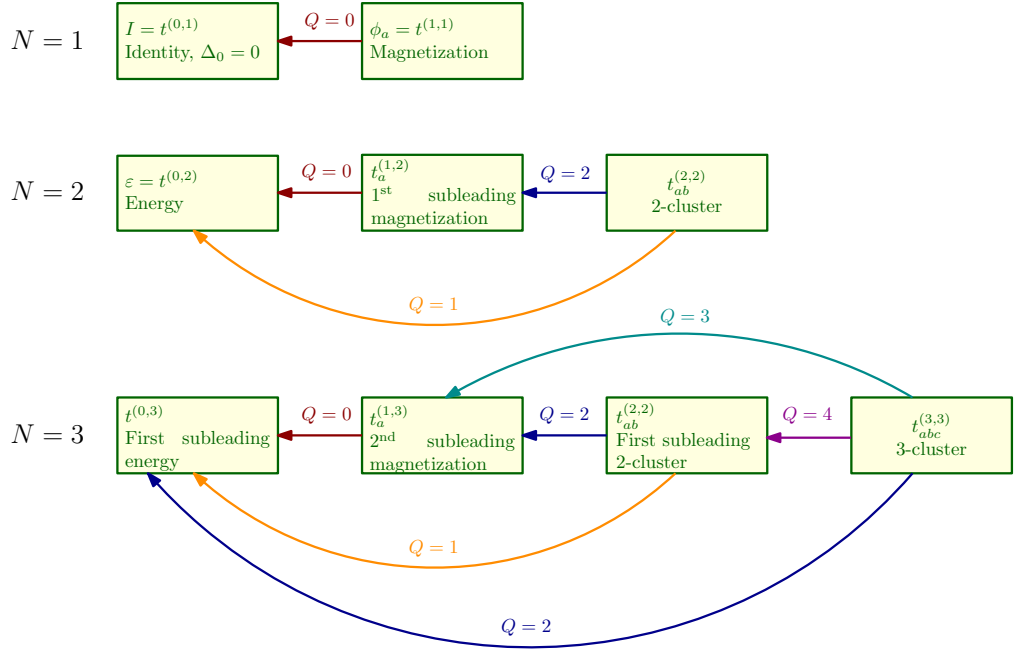}
\caption{General mixing structure of the scalar operators in the Potts model in arbitrary dimension. Arrows correspond to Jordan cells in the scale transformation generator, and are hence associated with geometrical logarithmic correlation functions. }
\label{FigMixing}
\end{figure}

In the previous sections, we have analyzed in details the mixing that occurs for $Q=0,1,2$ for operators that act on $N=1$ and $N=2$ spins. Using our extensive representation theory results and the poles of the correlation functions in section~\ref{sec:discrete}, one can actually obtain very general results regarding the logarithmic structure of the Potts model in arbitrary dimension. Our results are summarized for $N=1,2,3$ in Fig.~\ref{FigMixing}, and the general pattern for higher values of $N$ should be quite clear from this picture. Each arrow corresponds to a Jordan cell for the scale transformation generator, and is thus associated with a logarithmic field mixing the two fields. Note in particular that our representation theory analysis predicts infinitly many coincidences of critical exponents in arbitrary dimension, both for bulk and boundary exponents. For instance, at $Q=0$, we can infer from our results that the magnetization operator should be mixed with the identity, the first subleading magnetization operator should be mixed with the energy, and so on for all the following subleading operators. In particular, this implies the following identity for the (bulk) critical exponent 
\begin{equation}
Q=0 \ : \ \ \ \Delta^{(p)}_0 = \Delta^{(p)}_{\sigma}, \ p=0,1,\dots
\end{equation}
In two dimensions, this amounts to claim that $h_{p+1,1}=h_{\frac{p+1}{2},0}$, which is indeed correct. For percolation ($Q=1$), we have argued that the 2-cluster operator should be mixed with the energy field, and this works as well for all their subleading contributions. In terms of critical exponents, this yields
\begin{equation}
Q=1 \ : \ \ \ \Delta^{(p+1)}_0 = \Delta^{(p)}_{2}, \ p=0,1,\dots
\end{equation}
As a last example, for the Ising model ($Q=2$), we find that the 2-cluster operator is mixed with the subleading magnetization, and that the 3-cluster operator is mixed with the first subleading contribution to the energy. Once again, in terms of exponents this means that 
\begin{align}
Q=2 \ : \ \ \ \Delta^{(p+1)}_{\sigma} &= \Delta^{(p)}_{2}, \ p=0,1,\dots \\
              \Delta^{(p+2)}_{0} &= \Delta^{(p)}_{3}, \ p=0,1,\dots
\end{align}
We emphasize that these identities satisfied by the critical exponents are quite remarkable, as they should hold in any dimension. In $d=2$ they hold explicitly by the results of section~\ref{secCFTexact} and symmetries of the Kac table---or conversely, the fact that they {\em should} hold corroborates the identification of critical exponents made in section~\ref{secCFTexact}.

All these Jordan cells correspond to logarithmic correlation functions at the critical point which could be computed exactly in the same spirit that the many examples addressed in the previous sections. These logarithmic correlation functions can all be given a clear geometrical interpretation using the results of section~\ref{sec:discrete}.

\section{Conclusion}

In this paper, we have generalized the ideas of Cardy~\cite{CardyReplica} and used extensively the $S_Q$ symmetry of the Potts model to conjecture its (scalar) operator content in arbitrary dimension. Using representation theory, we have computed exactly the two- and three-point functions of these operators, and we have argued that logarithmic corrections appear at the critical point to cancel poles in correlation functions. We have analyzed the general structure of the scalar sector of logarithmic CFT underlying the critical point of the Potts model in arbitrary dimension, and a remarkable feature of our approach is that it provides a clear geometrical interpretation of all the correlation functions of the (logarithmic) operators of the theory. Our results hold both in the bulk or near a boundary, and imply remarkable identities that should be satisfied by (surface or bulk) critical exponents, in arbitrary dimension. 

It would be very interesting to extend our study to non-scalar operators, that is, operators with a non-vanishing spin that would transform non-trivially under rotations. This would allow us to construct for instance the stress-energy tensor, and analyze its logarithmic structure in arbitrary dimension. Our systematic study could also be generalized to the $O(n)$ model~\cite{CardyReplica} (see appendix~\ref{generalizationOn} for some ideas in that direction). The analysis of the disordered Potts model, for which the replicated theory has symmetry $S_Q \times S_M$, in the replica limit $M \to 0$, would also be extremely interesting.  

We stress again that we have analyzed the Potts model by using its $S_Q$ symmetry and---for the cases when it stands at a second-order phase transition---conformal invariance. The analysis
of any physical problem can be considered complete only when all its symmetries have been taken into account; but correct, albeit less complete results can be expected if only a subset of the
symmetries are used. It is known that in $d=2$ the Potts model transfer matrix (with free boundary conditions) commutes with the generators of the quantum algebra $SU(2)_q$ \cite{PS90}, where
$\sqrt{Q} = q + q^{-1}$, and a cognate symmetry (that extends to periodic boundary conditions) has been discussed in \cite{RS07sym}. These symmetries have not been taken into account here.
It follows that the specialization of our analysis to $d=2$ leads to correct results (as witnessed by the comparison, made in section~\ref{sec:Qto0}, of the case $Q \to 0$ to alternative approaches,
and by the numerical results of \cite{VJSPercoLogs} for $Q \to 1$); but those results are not complete. In particular, indecomposability and logarithmic correlation functions occur in $d=2$ whenever
$q$ is a root of unity \cite{PRZ,RS07,Vasseur_beta}, which includes cases when $Q$ is not a non-negative integer (such as $Q=4 \cos^2(\pi/5)$) that are not accounted for by our approach.
And for $Q$ a non-negative integer, there exist in $d=2$ Jordan cells that are not accounted for by our partial analysis. But the results obtained for the Jordan cells that are within the scope of our
approach are nevertheless correct.
Obviously we cannot exclude as well that other (yet unidentified) symmetries than those considered here may be operative for $d>2$, and possibly be specific to each choice of $d$.
In any case, it should be possibly to refine our results by extending the study to non-scalar operators.

\section*{Acknowledgments}

This work was supported by the French Agence Nationale pour la Recherche (ANR Projet 2010 Blanc SIMI 4: DIME), the Quantum Materials program of LBNL (RV), and the Institut Universitaire de France (JLJ). We warmly thank Hubert Saleur for collaboration on the related paper~\cite{VJSPercoLogs} which led to the present study. We also thank John Cardy, Raoul Santachiara and Jacopo Viti for discussions.

%
%
%
%
%
%
%
%

\newpage

\appendix

\section{Some percolation 3-point correlation functions on the lattice}
\label{app3pointPerco}

In this appendix, we gather some lattice 3-point functions in the percolation case ($Q=1$). Using the definitions of the operators $\psi_{ab}  (r_i) \equiv t^{(2,2)} (r_i) + \frac{2}{Q(Q-1)}  \varepsilon (r_i)$ and $\epsilon (r) \equiv  t^{(0,2)}(r) - \frac{Q-1}{Q} \Pdiff$, together with the results of section~\ref{secThreePointlattice}, one can readily obtain the following 3-point functions at $Q=1$
\begin{equation}
\langle \epsilon (r_1) \epsilon (r_2) \psi_{ab} (r_3)\rangle = 2 \left(\PPa -\PPb \right).
\end{equation}
\begin{multline}
\langle \epsilon (r_1) \psi_{ab} (r_2) \psi_{cd} (r_3)\rangle = \left(\delta_{ac}\delta_{bd}+\delta_{ad}\delta_{bc} +\delta_{ac}+\delta_{bd}+\delta_{ad}+\delta_{bc} \right) 2\left[\PPa -\PPb \right] \\
+ 4 \left( \PPcmm + \PPdmm - \Pdiff \Pa(r_{13}) + 2 \leftrightarrow 3 + 3 \ \PPb - 2 \ \PPa\right).
\end{multline}
\begin{multline}
\langle \psi_{ab} (r_1) \psi_{cd} (r_2) \psi_{ef} (r_3)\rangle =  \left[ 2 \PPa (\delta_{ac}\delta_{ce}\delta_{bd}\delta_{df} + \dots)
+2 \left(\PPa-\PPb \right) (\delta_{ac}\delta_{ce}\delta_{bd} + \dots) \right.\\
\left. - 2 \PPb (\delta_{ac}\delta_{ce} + \dots) + \PPb (\delta_{ac}\delta_{de}\delta_{fb} + \dots) +  \left(2 \PPa-\PPb \right)(\delta_{ac}\delta_{be} + \dots) 
\right.\\
\left.+ 4  \left(\delta_{ac}\delta_{bd}+\delta_{ad}\delta_{bc} +\delta_{ac}+\delta_{bd}+\delta_{ad}+\delta_{bc}\right) 
\right.\\
\left. 
\times\left(2\PPb-\PPa + \PPcmmm + \PPdmmm - \Pdiff \Pa(r_{12}) \right) + 3 \leftrightarrow 1 + 3 \leftrightarrow 2
\right.\\
\left.
+ 8  \left(\PPh +
\PPe +
\PPf +
\PPg -
\Pdiff \left[ \Pb(r_{12}) + \Pc(r_{12}) + \dots \right] + 2 \Pdiff^3 \right)
\right.\\
\left.
-8 \left(\PPc + \PPd - \Pdiff \left(\Pa(r_{12}) + \dots \right) -4 \PPb  \right)
\right].
\end{multline}

\section{A short overview of the generalization to the $\mathcal{O}(n)$ model}
\label{generalizationOn}

\begin{figure}[!t]
\centering
\includegraphics[width=0.8\textwidth]{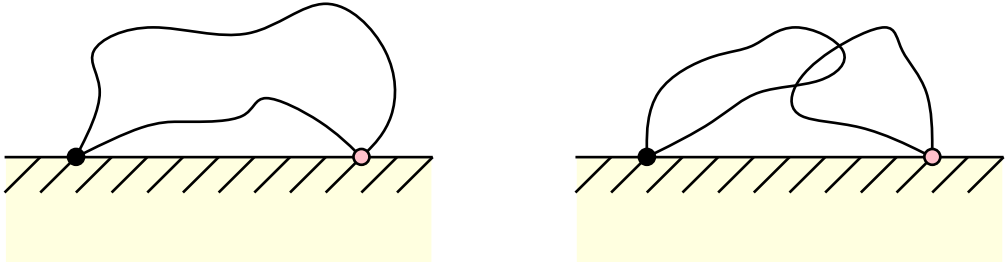}
\caption{Left: The number of $2$-leg watermelon configurations (two self-avoiding walks) starting from a fixed point on the boundary and coming back to another point (not fixed) on the boundary, with a total number of monomers $S$, scales as $Z_0(S) \sim {\rm e}^{T_c S} S^{\gamma-1}$~\cite{DuplantierSaleur}. Right: Configurations with two self-avoiding loops attached to the boundary, with a total number of monomers $S$, with the constraints that the first loop is attached to the origin, and that both loop must intersect at least once. This scales as $Z(S) \sim {\rm e}^{T_c S} S^{\gamma-1} \log S$~\cite{CardyReplica}. The ratio of these two quantities thus behaves as $Z(S)/Z_0(S) \sim \log S$.}
\label{FigLogsPoly}
\end{figure}

In this appendix, we briefly explain how our analysis of the Potts model
could be generalized to the case of $O(n)$ model in $d$ dimensions with action
\begin{equation}
S = \int {\rm d}^d x \left( \frac{1}{2} \sum_{a=1}^{n} (\partial_{\mu} \phi_a)^2+ \frac{m^2}{2} \sum_{a=1}^{n}  \phi^2_a + g \sum_{a,b} \phi^2_a \phi^2_b \right).
\end{equation}
We will focus on the field theory although a similar analysis
could be made at the level of a lattice Heisenberg Hamiltonian
of $n$-component spins $H=- J \sum_{\langle i,j\rangle} \vec{S_i}.\vec{S_j}$.
Using the representation theory of the $O(n)$ group, one can classify
the scaling operators of the model. For instance, it was argued by Cardy~\cite{CardyReplica}
that the energy operator and the generalization of the 2-leg operator in $d$ dimensions
could be expressed as
\begin{subequations}
\begin{eqnarray}
\varepsilon(r) &=& \sum_a :\phi_a^2: , \\
\varphi^{(2)}_{ab}(r) &=& :\phi_a \phi_b:-\frac{1}{n}\sum_c :\phi_c^2: ,
\end{eqnarray}
\end{subequations}
where $\varphi^{(2)}_{ab}(r)$ is a traceless symmetric tensor.
Using once again the global $O(n)$ symmetry of the field theory, one can write down 
the general expression of the two-point functions of these fields at the critical point~\cite{CardyReplica}
\begin{subequations}
\begin{eqnarray}
\langle \varepsilon(r) \varepsilon(0) \rangle &=& 2 n A(n)r^{-2 \Delta_{\epsilon}} , \\
\langle \varphi^{(2)}_{ab}(r) \varphi^{(2)}_{cd}(0) \rangle &=& \tilde{A}(n) \left( \delta_{ac} \delta_{bd}+ \delta_{ad} \delta_{bc} -  \frac{2}{n}\delta_{ab} \delta_{cd}\right)r^{-2 \Delta_{(2)}},
\end{eqnarray}
\end{subequations}
where $A(n)$ and $\tilde{A}(n)$ are regular function of $n$ with a finite non-zero limit at $n=0$.
The $n \rightarrow 0$ limit of these equations is singular, and this yields logarithms as discussed in 
the context of the Potts model. This was argued to have a nice geometrical interpretation in terms
of intersecting self-avoiding walks in Ref.~\cite{CardyReplica}. More precisely, taking properly the $n \to 0$ limit, one finds the following logarithmic correlation
\begin{equation}
\langle :\phi^2(r): :\phi^2(0): \rangle \sim \log r \times r^{-2 \Delta_{2}} \times f\left(\frac{r}{\xi} \right),
\end{equation}
where we have allowed for the system to be off-criticality, with correlation length $\xi \sim \left|T-T_c \right|^{-\nu}$, $e^{-T}$ being the monomer fugacity. Restricting for now to the boundary case, a more physical expression is obtained, as usual when dealing with polymers, by considering the Laplace transform so as to deal with polymers of fixed length 
\begin{equation}
\int_{\rm boundary} ({\rm d} r) \langle :\phi^2(r): :\phi^2(0): \rangle = \int_{0}^{\infty} {\rm d} S {\rm e}^{-T S} Z(S).
\end{equation}
In this equation, $Z(S)$ counts the configurations with two self-avoiding loops attached to the boundary, with a total number of monomers $S$, with the constraints that the first loop is attached to the origin, and that both loop must intersect at least once\footnote{This last constraint comes from the normal order.}. Inverting the Laplace transform, one finds the following asymptotic behavior at the critical point $T=T_c$~\cite{CardyReplica}
\begin{equation}
Z(S) \sim {\rm e}^{T_c S} S^{\gamma-1} \log S,
\end{equation}
with $\gamma = \nu \left(1-2 \Delta_{2}\right)$. The important point in that equation is the $\log S$ term, which differs from the counting of $2$-leg watermelon configurations that would scale as $Z_0(S) \sim {\rm e}^{T_c S} S^{\gamma-1}$~\cite{DuplantierSaleur} (see Fig.~\ref{FigLogsPoly}). It would be interesting to check this formula using exact enumerations. Note also that a similar formula can be established in the bulk.

This mixing can also be interpreted as critical exponents coinciding at $n=0$. At one-loop, the scaling dimension of the energy is well-known
\begin{equation}
\Delta_{\varepsilon} = 2 \left(\frac{d}{2}-1 \right) + \frac{n+2}{n+8} \epsilon + \mathcal{O}(\epsilon^2),
\end{equation}
with $\epsilon = 4-d$, whereas the dimension of $p$-leg watermelon operator can be obtained readily computing
Gaussian operator product expansions (see {\it e.g.}~\cite{CardyBook})
\begin{equation}
\Delta_{p} = p\left(\frac{d}{2}-1 \right)+ \frac{p(p-1)}{n+8} \epsilon + \mathcal{O}(\epsilon^2).
\end{equation}
For $n=0$, one has $\Delta_{2}=\Delta_{\varepsilon}$ as expected. Note that in $d=2$ dimensions, one has $\Delta_{2}= 2 h_{0,1} = \frac{2}{3}$ and $\Delta_{\varepsilon}= 2 h_{1,2} = \frac{2}{3}$.

This line of thought can be generalized to more complicated cases. For example, we
find that the 3-leg operator reads
\begin{equation}
\varphi^{(3)}_{abc} = \phi_a \phi_b \phi_c - \frac{1}{n+2} \sum_d \phi_d^2 \left(\delta_{ab} \phi_c +\delta_{ac} \phi_b +\delta_{bc} \phi_a \right).
\end{equation}
One can see that some mixing should be expected at $n=-2$ between $\varphi^{(3)}_{abc} $ and the operator $\sum_b :\phi_b^2 \phi_a: $.
The latter has a form energy $\times$ magnetization so we expect it to be the first subleading magnetization operator.
Its one loop scaling dimension can be easily computed
\begin{equation}
\Delta^{(1)}_{\phi} = 3\left(\frac{d}{2}-1 \right) + \epsilon + \mathcal{O}(\epsilon^2),
\end{equation}
and we indeed find that $\Delta^{(1)}_{\phi}= \Delta_{3}$ for $n=-2$. 
It would be very interesting to study extensively the $O(n)$ model along those lines, as we did for the Potts model
in this paper. A closer look at the lattice model would then be necessary in order to interpret these correlation 
functions geometrically.

\end{document}